\documentstyle{mn}
\input epsf.sty

\def\Rg{R_{\rm g}}

\def\NH{N_{\rm H}}
\def\Ec{E_{\rm c}}
\def\Rin{R_{\rm in}}
\def\Din{D_{\rm in}}
\def\Rout{R_{\rm out}}
\def\Rst{R_{\rm unst}}

\def\ginga{{\it Ginga}\ }

\def\LEdd{L_{\rm Edd}}

\def\kT{k T_{\rm e}}
\def\mdot{\dot m}

\def\LX{L_{\rm X}}

\def\Firr{F_{\rm irr}}
\def\relrepr{{\tt relrepr}\ }
\def\MSun{{\rm M}_{\odot}}
\def\gs{{GS~2023+338}\ }
\def\taues{\tau_{\rm es}}
\def\tauabs{\tau_{\rm abs}}

\def\dof{{\rm dof}}
\def\Or{\Omega_{\rm r}}
\def\Tr{k T_{\rm r}}
\def\Tsoft{k T_{\rm soft}}
\def\taur{\tau_{\rm r}}
\def\RT{R_{\rm tidal}}
\def\Rcirc{R_{\rm circ}}
\def\Mdisc{M_{\rm disc}}

\def\Teff{T_{\rm eff}}
\def\DMC{\Delta M_{\rm c}}
\def\DMX{\Delta M_{\rm X}}

\title[Outburst of GS 2023+338 (V404 Cyg)]
{The May 1989 outburst of the Soft X-ray Transient GS~2023+338 (V404 Cyg)}
\author[P.T. \.{Z}ycki, C. Done and D.A. Smith]
    {Piotr T. \.{Z}ycki$^{1,3}$, Chris Done$^1$ and 
     David A. Smith$^2$\thanks{Present address: NASA/Goddard Space 
     Flight Center, Greenbelt, MD 20771, USA}\\ 
        $^1$ University of Durham, Department of Physics, 
       South Road, Durham DH1 3LE; chris.done@durham.ac.uk \\
        $^2$ Department of Physics and Astronomy, University of Leicester,
            University Road, Leicester LE1 7RH \\
        $^3$ Nicolaus Copernicus Astronomical Center, Bartycka 18, 
            00-716 Warsaw, Poland; ptz@camk.edu.pl}
 
\date{22 April 1999}
 
\pagerange{\pageref{firstpage}--\pageref{lastpage}}
\pubyear{1999}

\begin{document}

\maketitle
 
\begin{abstract}

We re-analyze archival {\it Ginga\/} data of the soft X-ray  transient 
source GS~2023+338 covering the beginning of its May 1989 outburst.
The source showed a number of rather unusual features: very high and 
apparently saturated
luminosity, dramatic flux and spectral variability (often on $\sim 1\,$sec time
scale), generally very hard spectrum, with no obvious soft thermal component
characteristic for soft/high state. 

We describe the spectrum obtained at the
maximum of flux and we demonstrate that it is very different from spectra
of other soft X-ray transients at similar luminosity. 
We confirm previous suggestions that
the dramatic variability was due to heavy and strongly variable 
photo--electric absorption. We also demonstrate that for a short time the
source's spectrum did look like a typical soft/high state spectrum but
that this coincided with very heavy absorption.

\end{abstract}
 
\label{firstpage}
 
 
\section{Introduction}

Low mass X--ray binaries, where a stellar mass black hole or neutron star
accretes matter from its Roche Lobe filling companion, are generally {\it
transient\/} systems. The mass accretion rate is low enough for the
accretion disc to be unstable, leading to long quiescence periods followed by
dramatic outbursts (e.g.\ King et al.\ 1997a; King, Kolb \& 
Szuszkiewicz 1997b). For the
black hole systems, the outburst often shows a rapid rise from a very faint
quiescent state, reaching luminosities close to the Eddington limit in a course
of a few days.  The outburst then declines roughly exponentially, with a
characteristic time scale of 30--40 days in both the X--ray and optical flux
(e.g.\ Chen, Schrader \& Livio 1997).
During the decline the X--ray spectra and variability go through a sequence of
well defined states. For luminosities close to the Eddington limit, 
$\LEdd$, the
spectrum has both a strong soft component from the accretion disc and a strong
power law tail (the very high state). At lower luminosities the hard tail
decreases and steepens, so the spectrum is dominated by the soft component (the
high state). There is then a dramatic transition, where the soft component
decreases substantially in temperature and luminosity, and the spectrum is
instead dominated by a hard power law tail (the low state) (see e.g.\ 
Tanaka \& Lewin 1995; van der Klis 1995; Tanaka \& Shibazaki 1996).

\gs broke with this pattern in several ways.  
Firstly, it never clearly showed
the very high or high state spectrum even though its luminosity was close to 
the
Eddington limit. Secondly, it showed dramatic flux and spectral variability,
part of which can be attributed to a heavy and strongly variable 
photo--electric
absorption (Tanaka \& Lewin 1995; Oosterbroek et al.\ 1997). Thirdly, its
optical spectrum was unlike that from any other transient system, with strong,
broad lines from H, HeI and HeII: normally these lines are rather weak (Casares
et al.\ 1991).

In this paper we analyze in detail archival \ginga data of \gs covering the
beginning of its outburst, 23--30 May 1989, attempting to deconvolve the 
primary
spectrum from effects of subsequent reprocessing (absorption, Compton
reflection, line fluorescence).  The spectrum at the peak luminosity is very
different to that seen in more canonical transient systems such as Nova
Muscae 1991. We postulate that this spectrum represents super Eddington 
accretion rates (Inoue 1993), 
and speculate that this and many of the other peculiar properties of \gs
are due to the wide separation of the binary, as suggested for GRO J1655--40
(Hynes et al.\ 1998). The accumulating accretion disc is
then very large, so that when the disc instability is triggered there is much
more mass in the quiescent disc than in a more typical transient. The accretion
can then be super Eddington, giving rise to some form of strong mass outflow
which manifests itself as heavy photo--electric absorption, and produces the
intense optical line emission. This absorption shrouds the source as the
accretion rate declines below Eddington, masking its transition to the standard
very high spectrum expected at these luminosities.  As the
outflow expands with time, the absorption becomes less extreme. Eventually,
there are times when the intrinsic source spectrum is not obscured, but by this
point the mass accretion rate has declined so that the source is in the 
low/hard state.

\section{Data reduction and background subtraction}
 
Data were extracted from the UK \ginga data archive at Leicester University
and reduced in the usual manner. Background subtraction poses a problem for
a source as bright and  hard as \gs since the background monitors 
(most importantly the Surplus above Upper Discriminator, or SUD) are 
contaminated
by counts from the source. We have used the same method as described in
\.{Z}ycki, Done \& Smith (1999; hereafter Paper I) to recover the 
uncontaminated SUD values and then we used the
'universal' background subtraction method (Hayashida et al.\ 1989) to
estimate and subtract the background. We assume 0.5 per cent systematic errors
in the data unless stated otherwise.

\begin{figure}
 \epsfxsize=0.5\textwidth
 \epsfbox[10 190 590 670]{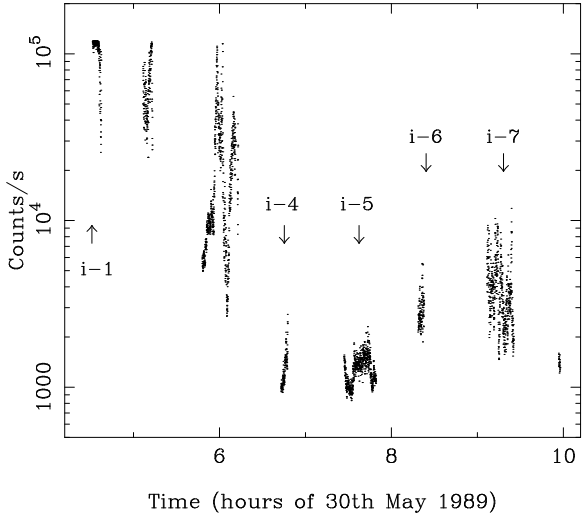}
 \caption{{\it Ginga\/} light--curve (count rate in 1--30 keV; 2 sec time bins)
 on 30th May, when the source's flux reached its maximum and apparently 
 saturated as well as showed dramatic
variability. The count rate was corrected for background, dead-time and
aspect. Beginning of good data, $t_0$, is 4:31:13.
Data for the unabsorbed spectrum (Section~\ref{sec:may30}; 
Figure~\ref{fig:may30spec}; Table~\ref{tab:may30spec}),
and for the absorbed one shown in Figure~\ref{fig:s329_abs} were extracted
from time interval $i-1$. Spectra from time intervals $i-4$ and $i-5$ 
are analyzed in Section~\ref{sec:soft}. They are consistent with soft/high
state spectra of SXT, when corrected for absorption.
\label{fig:may30_lc}}
\end{figure}

\begin{figure}
 \epsfxsize=0.5\textwidth
 \epsfbox[10 180 600 690]{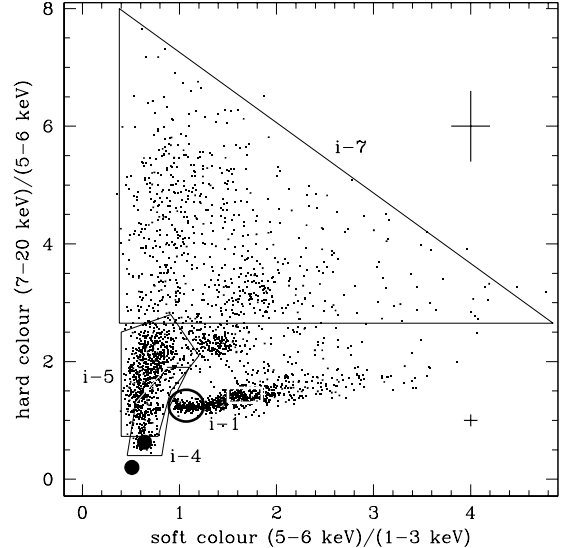}
\caption{
Colour--colour plot for data obtained on 30th May. Time bin is 1 second.
The circle marks the region where data for the unabsorbed spectrum were
extracted from (Section~\ref{sec:may30}), while the rectangle marks similar
region for the absorbed spectrum shown in Figure~\ref{fig:s329_abs}.
Contours labeled $i-4$ and $i-5$ mark approximate regions where the soft 
state data were extracted from (cf.\ Figure~\ref{fig:may30_lc}; 
Section~\ref{sec:soft}). Very rapid and chaotic variability was observed
during time interval $i-7$, most likely due to photo-electric absorption 
(Section~\ref{sec:heavyabs}). For comparison, the two big dots show position
of Nova Muscae 1991 on January 11 and January 16 (see Section~\ref{sec:compar},
Figure~\ref{fig:gs_nm}). The crosses show typical error bars on the colours:
the bigger one is for $i-7$, the smaller one for earlier observations.
\label{fig:colcol}}
\end{figure}

\section{Models}

We use a variety of models for spectral description.  For the soft component we
generally use the multi-temperature blackbody model (Mitsuda et al.\ 1984). For
simple estimates we use the simple version implemented as {\tt diskbb} in {\sc
XSPEC}. This however does not include the colour temperature correction (e.g.\
Shimura \& Takahara 1995), or the torque-free boundary condition term in the
expression for temperature, $1-\sqrt{6\Rg/r}$ (assuming Newtonian dynamics;
Shakura \& Sunyaev 1973), where $\Rg\equiv G M/c^2$. We implemented these
corrections in a model called {\tt diskspec} (see also Gierli\'{n}ski et al.\ 
1998). The main parameter can be chosen to be either the temperature at the
inner disc radius or the ratio of mass accretion rate and the mass of the
central object.

\begin{table*}
 \caption{Model fitting of the May 30.\ unabsorbed spectrum.
\label{tab:may30spec}}
   \begin{tabular}{lccccc}
 parameter & units  &  0   &         A    &    
                                         B    &  
                                                C \\
\hline
$\NH$  & $10^{22}\,{\rm cm^{-2}}$    &
                       3.7  &  $0.50^{+0.24}$   & 
				$0.50^{+0.13}$ & 
					$0.50^{+0.22}$     \\
$\Tsoft$ &     keV      & 
                       0.23  &  $0.270\pm 0.015$ &  
				$0.322\pm 0.015$       & 
					$0.281\pm 0.013$   \\
$\Din$    &   $\Rg$    & 
                      370 &  $71^{+15}_{-12}$ & 
				$41 \pm 8$      &  
					$62^{+13}_{-10}$   \\
$\Gamma$  &    & 
                     1.57 &  $1.03\pm 0.04$   & 
				$1.690^{+0.023}_{-0.011}\,^{a\,b}$&
					$1.700^{+0.013}_{-0.011}\,^{a\,b}$ \\
$E_c$    &  keV       &  
                      60  &  $19.6^{+1.2}_{-1.0}$&
				--              &
                                            --             \\           
$\kT$    & keV     &  -- &  --   & $9.2^{+0.4}_{-0.2}\,^b$ &  
                                               $9.6^{+0.7}_{-0.5}\,^b$  \\
$\taues$ &         &  -- &  --   & -- &  
                                               $6.49\pm 0.07\,^b$       \\
$T_0$    & keV     &  -- &  --   & $0.83\pm 0.03$ &  
                                               $1.44\pm 0.09$       \\
\hline
$\Or$    &         &  -- &  --   & $0.13^{+0.05}_{-0.03}$ & 
					$0.17^{+0.16}_{-0.05}$ \\
$\xi$    & erg/cm s &-- &  --   & $(2.5^{+7.5}_{-2.0})\times 10^3$ &
					$(6^{+18}_{-4})\times 10^3 $ \\
$\Rin$   & $\Rg$   &  -- &  --   & $6^{+1.1}$  & 
					$6^{+2}$  \\
$E_{\rm edge}$& keV&  -- & $7.70\pm 0.13$ & -- & -- \\
$\tau_{\rm edge}$& &  -- & $1.19\pm 0.13$ & -- & -- \\
EW       & eV      &  -- & $80^{+13}_{-20}$ & $67 \pm 16$ & 
					$70 \pm 17 $ \\
$\chi^2/$dof &    & 555/32 &  23.5/29 & 37.3/27 &
                                       25.7/27 \\
 \hline 
   \end{tabular}
\medskip

Model 0: absorption*(disc blackbody + cutoff power law)

Model A: absorption*smeared edge*( disc blackbody + cutoff power law + 
gaussian) 

Model B: absorption*(disc blackbody + comptonized blackbody ({\tt thComp}) + 
\relrepr + gaussian)

Model C: absorption*(disc blackbody + comptonized disc blackbody 
({\tt thComp}) 
+ \relrepr +gaussian) 


$\NH$ -- hydrogen column density. 
Its hard lower limit (interstellar value) is assumed 0.5.\\
$\Din$ -- inner disc radius computed from the amplitude of the 
      {\tt diskbb} model \\
$\kT$, $\taues$ -- electron temperature and optical depth of the 
 comptonizing cloud \\
$T_0$ -- temperature of the seed photon input spectrum for comptonization \\
$E_c$ -- e-folding energy in the cutoff power law model \\
$\Or,\ \xi$ -- amplitude and ionization parameter of the reprocessed
component \\
$\Rin$ -- inner disc radius determining the level of relativistic smearing \\
EW --  equivalent width of the additional gaussian line \\
$^{a}\,$ Asymptotic value of $\Gamma$ in the ST80 solution. \\
$^b\,$ $\Gamma$, $\kT$ and $\taues$ are of course not independent. \\
Parameters' uncertainties are 90 per cent confidence limits for one
interesting parameter i.e.\ $\Delta\chi^2=2.71$.
\end{table*}

We use an analytic thermal comptonization model to describe the hard component:
{\tt thComp}, based on
solution of the Kompaneets equation (Lightman \& Zdziarski 1987).  For accurate
modelling of inverse Compton spectra we also use a Monte Carlo simulation
code. It is based on standard methods of simulations of the inverse Compton
process as described in detail by Pozdnyakov, Sobol \& Sunyaev (1983) and
G\'{o}recki \& Wilczewski (1984) (see Appendix A).
 
The X--ray reprocessed component is modelled using the angle-dependent
Green's functions of Magdziarz \& Zdziarski (1995) convolved with a given
continuum model to produce the reflected continuum, with photo--electric
opacities calculated by a simple photo-ionization code described in 
Done et al.\
(1992). For the iron fluorescent/recombination K$\alpha$ line we used the Monte
Carlo code of \.{Z}ycki \& Czerny (1994). We constructed Green's functions for
the line emission (i.e.\ emission for a monochromatic irradiation flux), as
functions of ionization and iron abundance, which can be folded with a given
continuum model, to compute total line profile, which is then added to the
reflected continuum. Parameters of the model are: the amplitude of the total
reprocessed component, defined as the solid angle of the reprocessor as seen
from the X--ray source, $\Omega$, normalized to $2 \pi$, $\Or\equiv \Omega/2
\pi$ and the ionization parameter, $\xi \equiv 4\pi F_{\rm X}/n$, where $F_{\rm
X}$ is the illuminating flux in the 5 eV -- 20 keV band and $n$ is the electron
number density.  We assume cosmic abundances of Morrison \& McCammon (1983) 
with the possibility of a variable iron abundance in the reprocessor. 
For more model details see Paper I.

The total reprocessed spectrum can then be convolved with
the {\sc XSPEC} {\tt diskline} model (Fabian et al.\ 1989; modified to include
light bending in Schwarzschild metric) to simulate the relativistic 
smearing. The main parameter here is the inner radius of the disc, $\Rin$,
assuming a given form of the irradiation emissivity, for which we adopt
$\Firr(r) \propto r^{-3}$ (see Paper I). The outer disc radius is fixed
at $10^4\,\Rg$. This is smaller than the expected outer radius of the disc
in \gs ($\sim 10^5\,\Rg$; see Section~\ref{sec:discuss}), but with our
assumed $\Firr(r)$ the fractional contribution from the ring 
$10^4$--$10^5\,\Rg$ would be negligible, $\approx 5\times 10^{-4}$.
The total model will be referred to as {\tt relrepr}.

Strong photo--electric absorption is modelled by either the  simple
formula $F \propto \exp(-\tau_{\rm eff})$ where
$\tau_{\rm eff} = \sqrt{\tauabs (\tauabs + \taues)}$ (Rybicki \& Lightman 1979;
called {\tt thabs})
or by a proper Monte Carlo transmission code for spherical geometry.

We fix the inclination of the source at $i=56^{\circ}$ and assume the mass of
the black hole is $12\,\MSun$ (Shahbaz et al.\ 1994). We assume the 
interstellar
$\NH=5\times 10^{21}\,{\rm cm^{-2}}$, corresponding to $A_V=3$ (Chen et al.\
1997; Shahbaz et al.\ 1994).
 
The spectral analysis was performed using {\sc XSPEC} ver.\ 10 (Arnaud 1996)
into which all the non-standard models mentioned above were implemented as
local models. 
 
\section{The unabsorbed spectrum}

\label{sec:may30}

On the 30th of May \ginga observed \gs in a very bright state
(see Fig.~\ref{fig:may30_lc}). The observed source flux reached $6.5\times 10^{-7}$ 
erg/s/cm$^2$ (1 -- 40 keV; corresponding to 
$\LX\approx 10^{39} {\rm erg\ s}^{-1}$ at $d=3.5$ kpc)
and it apparently saturated at this level. 
For about 200 sec after the beginning of observation the source spectrum 
was very stable (Fig.~\ref{fig:colcol}) and it did not show any strong 
photo--electric absorption, so we used the 200 sec average spectrum
for detailed analysis.

\subsection{Phenomenological description}

Generally, the data can be described (Table~\ref{tab:may30spec}) as a sum of a
soft component and a hard cutoff power law component ($\Gamma\approx 1.0$) with
e-folding energy of $\Ec\approx 20\,$keV (i.e.\ the cutoff is obviously visible
in the \ginga band). Superimposed on this are spectral features near 6--10
keV. These features can be phenomenologically modelled by a narrow gaussian at
6.4 keV (EW=77 eV) and the smeared edge (Ebisawa 1991) with $E\approx 7.7\,$keV
and $\tau\approx 1.2$ (Model A in Table~\ref{tab:may30spec}.  The presence of
the spectral features is highly significant; the best fit model using only two
continuum components gives $\chi^2=555/32\,$dof (Model 0 in
Table~\ref{tab:may30spec}). 

The soft component can be described equally well by both simple blackbody and
disc blackbody model. This is due to the fact that only the Wien cutoff is
visible in \ginga band.  Assuming the model which would give the lowest
bolometric correction i.e.\ a blackbody spectrum and assuming further that at
that time there was no photo-electric absorption in excess of the interstellar
value, the derived temperature is $\Tsoft=0.21\,$keV and the normalization of
the component corresponds to the emitting area of $\approx 3\times 10^{6}\,{\rm
km^2}\approx (90\,\Rg)^2$. Extrapolated to lower energies, this component then
contributes significantly to the total energy budget, with a total bolometric
flux of $10^{-6}$ ergs cm$^{-2}$ s$^{-1}$. Thus the lowest possible luminosity
is around the Eddington limit, but it is likely to be rather higher since 
the soft
spectrum must be broader than a single blackbody and the absorption may be
rather higher than the (poorly known) interstellar value of $5\times 10^{21}$
cm$^{-2}$.

\begin{figure}
\epsfxsize = 0.45\textwidth
\epsfbox[0 190 550 720]{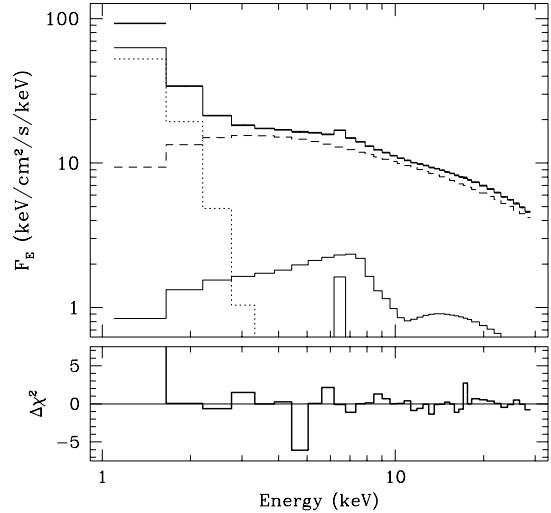}
\caption{Source spectrum on 30 May, 4:31:13 -- 4:34:33 (200 sec), when
the flux apparently saturated at very high level (Figure~\ref{fig:may30_lc},
part of time interval $i-1$). The spectrum can be 
described as optically thick comptonization of a disc blackbody seed
spectrum of temperature $k T_0\approx 1.4\,$keV, electron temperature
$\kT\approx 10\,$keV and $\taues\approx 6$, with a corresponding 
reprocessed component (model C in Table~\ref{tab:may30spec}).
An additional soft component is also required: its
temperature is rather low: $k T \approx 0.28\,$keV and its amplitude
corresponds to inner radius of the emitting  disc $\sim 60\,\Rg$
(for the {\tt diskbb} model). Counts
in the first channel are much above the model contribution suggesting more
complicated modelling is necessary. This channel was ignored in spectral 
fitting. \label{fig:may30spec}}
\end{figure}

\subsection{Analytical comptonization models}

A cutoff power law is not necessarily a good approximation to a comptonized
spectrum. We first use the {\tt thComp} model as an analytic description of the
spectrum, including its self--consistently calculated reprocessed spectrum. 
We also include a separate Fe K$\alpha$ line from fluorescence from distant
material. The presence of this was inferred by Oosterbroek et al.\ (1996) 
from timing analysis of the May 30th data. Since it is from distant material 
we fix its energy at 6.4 keV, and assume that it is narrow. We also 
include a narrow K$\beta$ line at 7.05 keV, tied to 11 per cent of the 
intensity of the narrow K$\alpha$ line. The best fit {\tt thComp} model has 
$\chi^2=37.3/27\,$dof,
with low electron temperature, $\kT = 9.2\pm 0.4\,$keV, rather large
optical depth, $\taues = 6.73\pm 0.07$, but requires that the seed photon 
blackbody temperature is very much higher than that of the observed soft flux,
with $T_0 = 0.83\pm 0.04\,$keV. Fixing the seed photons to the temperature of
the observed soft excess gives a very much poorer fit as it is unable to
reproduce the curvature seen in the hard spectrum at low energies.
This best fit model also requires a strongly ionized and smeared reprocessed 
component, with amplitude $\Or\approx 0.13$ (Model B in Table~\ref{tab:may30spec})

A significantly better fit can be obtained assuming a broader spectral 
distribution of the seed photons for comptonization. We have constructed 
a version of the {\tt thComp} model in which the input soft 
photon spectrum is a disc blackbody, rather than a simple blackbody.
This model (model C in Table~\ref{tab:may30spec}), with its corresponding 
reprocessed component, can now adequately describe the data, with
best fit $\chi^2=25.7/27\,$dof (i.e. $\Delta\chi^2=11.6$ compared to 
previous case).
The comptonizing cloud is rather similar to that derived previously. It is 
optically thick, $\taues= 6.49\pm 0.07$
and has temperature of $\kT = 9.6^{+0.7}_{-0.5}\,$keV.
The amplitude of the reprocessed component is much smaller than 1, 
$\Or = 0.17^{+0.16}_{-0.05}$, but the reflector properties are
nevertheless well constrained. It has to be strongly ionized 
($\xi\approx 6000$, so H-like iron ions dominate) 
and strongly smeared, $\Rin=6\,\Rg$.
The narrow gaussian at 6.4 keV is still required, its equivalent width
is $\sim 70\,$eV.
The best fit model spectrum is plotted in Figure~\ref{fig:may30spec}.

An almost equally good fit ($\chi^2=29/27$) can be obtained by replacing the
gaussian lines by an unsmeared, neutral reflected spectrum, so it is not
possible to distinguish from the spectrum whether the distant material inferred
from the timing analysis (Oosterbroek et al.\ 1996) is optically thick or 
thin for electron scattering.

Hard lower limit on $\NH$ equal to the likely interstellar value of 
$5\times 10^{21}\,{\rm cm^{-2}}$ was imposed in the above fits. Allowing
$\NH$ to be a free parameter in model C we actually find that the best fit 
value is 0,
although the fit is now better by only $\Delta\chi^2=2.6$ (F-test significance
$\approx 90$ per cent). This may indicate possible complexity of the soft
component with further evidence for it coming from the observed excess
of counts in the first channel (Figure~\ref{fig:may30spec}). We note however
that the excess cannot be solely the result of (unlikely) changes 
of $\NH$ during the integration interval, since the excess remains 
(although reduced) even if $\NH$ is fixed at 0.

\subsection{Monte Carlo comptonization models}

The analytical solutions for comptonization Green's functions of 
Sunyaev \& Titarchuk (1980), Lightman \& Zdziarski (1987) and Titarchuk (1994)
are not accurate
when the seed photon energy is close to the plasma temperature, even in the
diffusion approximation. Discrepancies also appear close to the plasma
temperature (irrespectively of the seed photons energy) as the models 
are usually not able to correctly reproduce the shape of the high energy 
cutoff/Wien peak (see e.g.\ Fig.~2 in Titarchuk 1994). Since in our case the 
considered energy range is close to both the seed photons energy and plasma
temperature, we can expect differences between analytic and simulated spectra. 

\begin{figure}
\epsfxsize = 0.45\textwidth
\epsfbox[30 300 550 690]{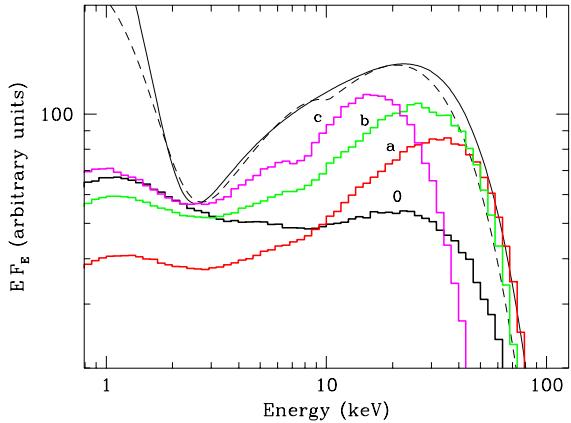}
\caption{Comptonization spectra in the geometry where the source of seed 
photons
is external to the comptonizing cloud {\it cannot\/} explain the unabsorbed 
spectrum of GS~2023+338. Solid and dashed curves
are best fit models C and 0 (Table~\ref{tab:may30spec}), respectively.
The seed photons spectrum was assumed a disc blackbody of temperature 
$T_0=0.28\,$keV,
as observed. Spectrum 0 is for $\kT=11\,$keV and $\taues=6$, i.e.\ the same
parameters as model C with central source, but with the external illumination
the spectrum is much softer. Spectra {\it a}, {\it b}  and {\it c\/}
are for $\kT=11,\ 9,\ 5$ keV and $\taues=8,\ 9,\ 15$, respectively,
and are normalized so that they do not exceed the observed spectrum.
\label{fig:ext_spec}}
\end{figure}

However, we demonstrate in Appendix A that for these parameters it is 
possible to  approximate 'exact' spectra (obtained by Monte Carlo 
simulations) to better than 1 per cent by the analytical models.
The results obtained in previous section are therefore vindicated
even though we are not able to do proper fitting of Monte Carlo spectra
to the data. The spectral curvature observed in the hard component 
is not an artifact of the deficiencies of analytic comptonization models. The
Monte  Carlo spectra also require that the seed photons are at much higher
temperatures than those from the observed soft excess. 

The analytic and Monte  Carlo models compared here both assume that the seed
photons are distributed within the comptonizing cloud. However, this is not the
case if the hard X--rays form a central source, surrounded by the accretion 
disc
(see e.g.\ the review by Poutanen 1998 for observational pointers to such a
geometry). We computed Monte Carlo spectra assuming a central, spherical plasma
cloud of radial optical thickness $\taues$ and electron temperature $\kT$,
surrounded by a geometrically thin disc. The disc is a source of soft photons
with a multi-temperature (disc) blackbody spectral distribution parameterized 
by
the temperature at the inner edge, $T_0$. These soft photons then form both the
observed soft excess spectrum and act as the {\it external\/} illuminating seed
photons for the comptonizing cloud.  Figure~\ref{fig:ext_spec} shows examples 
of
inverse Compton spectra obtained in such a geometry. It is not possible to
generate a hard and broad enough spectrum in such a geometry, given the
constraints on the electron temperature from the observed high energy roll-over
and the temperature of the observed soft component.
To produce a sufficiently hard spectrum, the optical depth of the cloud has
to be large enough for the comptonized spectrum to resemble a Wien peak,
which is then rather narrower than the observed spectrum.

Thus it seems inescapable that the comptonized spectrum seen is {\it not\/} 
produced by scattering the observed soft photons, but rather has as its seed
photons a rather higher temperature component which cannot be seen directly
(probably because it is embedded in the optically thick comptonizing cloud).

\subsection{Comparison with Nova Muscae 1991}

\label{sec:compar}
\begin{figure*}
\epsfysize = 7 cm
\epsfbox[18 460 590 690]{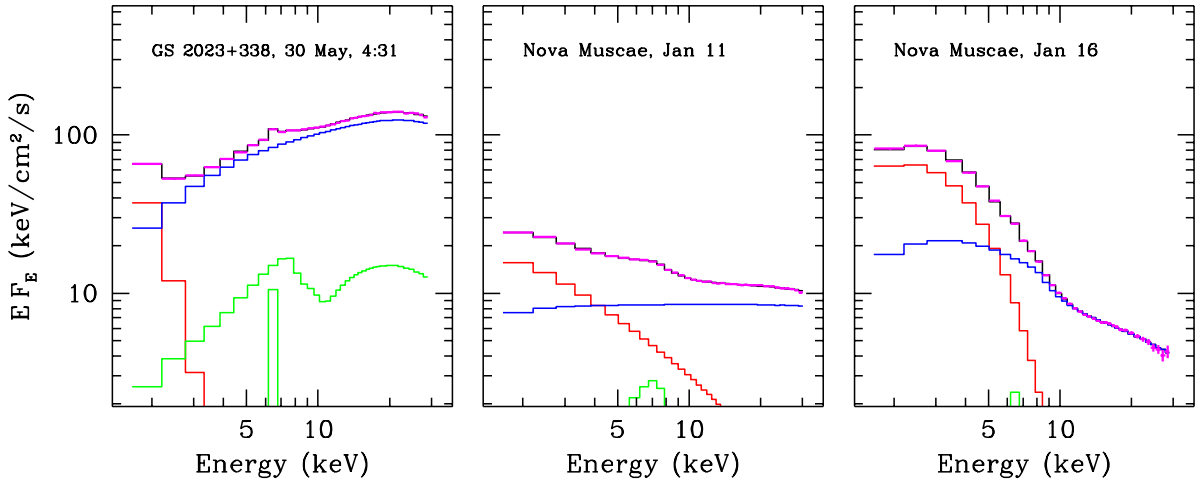}
\caption{Comparison of energy spectra of \gs and Nova Muscae 1991 at peaks of
their outbursts. The spectra (Jan 16 of Nova Muscae and the \gs one) are
very different even though sources' luminosities (in Eddington units) are
very similar and very close to 1. On Jan 11, when $L \approx 0.5\,\LEdd$,
the soft component cannot be described by a (disc) blackbody -- an additional 
power law tail is required. The strength of the soft component was comparable 
to that of the hard
component. Further increase in $\mdot$ results in weakening of the hard
component (its slope increases), and strengthening of the soft component
which now {\it can\/} be described by the disc blackbody model.
\label{fig:gs_nm}}
\end{figure*}

The spectral and timing behaviour of GS~2023+338 were quite different from
those of Nova Muscae 1991 at the peak of its outburst. Takizawa et al.\ (1997)
examined {\it Ginga\/} data of Nova Muscae 1991 covering first month
of its outburst, when the source luminosity was above $\sim 0.5\LEdd$.
We computed positions of the peak spectrum of 
\gs on their colour--colour and colour--count rate
diagrams (their Figure~9bc). The positions are well outside their diagrams,
in the directions indicating that the spectrum of \gs is much harder than
any of the spectra of Nova Muscae below $\sim 10\,$keV, but it is softer
than those above this energy. We illustrate this point in 
Figure~\ref{fig:gs_nm}, where we plot the spectra of both objects: 
two spectra of Nova Muscae are plotted: one obtained on 11 Jan 
($L\approx 0.5 \LEdd$) and the other  on 16 Jan, which has $L\approx \LEdd$ 
(number 1 and 6 in Takizawa et al.\ 1997; 
where the outburst of Nova Muscae was first detected on 8 Jan). 
The latter spectrum has similar luminosity to the 
peak spectrum of GS~2023+338, yet is very different in shape. For Nova Muscae
the soft component is much hotter ($T_0=0.87\pm 0.03\,$keV) and dominates
the luminosity, while the power law is very soft, 
$\Gamma = 2.76^{+0.08}_{-0.22}$. Even on Jan 11, where the spectrum
of Nova Muscae has a strong hard component, its spectrum is much softer than
that seen in GS~2023+338 at its peak luminosity. The peak GS~2023+338 spectrum
looks similar to the low state spectra seen in both Nova Muscae, GS~2023+338,
and other GBH at $L\approx 0.05 \LEdd$ except for its roll-over at $\sim 20$
keV, and much stronger soft component.

\begin{figure}
\epsfxsize = 0.45\textwidth
\epsfbox[30 200 550 690]{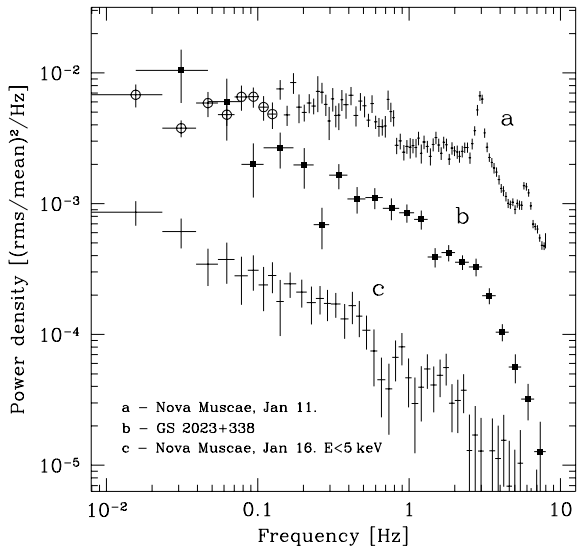}
\caption{Comparison of power spectra of Nova Muscae 1991 and GS~2023+338
at peaks of their outbursts. The same data as for energy spectra 
(Figure~\ref{fig:gs_nm}) are used. Only data between 1--5 keV are used
for the 16.\ Jan PSD of Nova Muscae.
\label{fig:i-1psd}}
\end{figure}

The power spectral density (PSD) of \gs during the first 200~s of the $i-1$
interval (excluding the dramatic intensity drop; Figure~\ref{fig:i-1psd}) has
intermediate properties between PSDs seen for the two high luminosity spectra 
of
Nova Muscae shown in Figure~\ref{fig:gs_nm} (cf. Figure~2 in Takizawa et al.\
1997, panels 1 and 6). Its shape is close to a power law, which in Nova Muscae
seems to correlate with energy spectra being dominated by a thermal component.
Its amplitude is however larger, comparable to the amplitude of the
(flat-topped) PSD obtained when the energy spectra show strong comptonized 
power
law. This PSD is nothing like that obtained for GBH low state spectra, where
the normalized variability amplitude is much larger, at $\sim 0.1$ at 1~Hz.
So even though the energy spectrum bears a (rather small) resemblance to 
the low
state spectrum, its variability behaviour clearly shows that it is very 
different.

Both the spectrum and PSD
of GS~2023+338 at its peak luminosity look very different from any known
spectral state of GBH.

\subsection{Discussion}

The spectrum and variability of GS~2023+338 at its peak luminosity show that
this is very different from any known GBH spectral state.  The soft component
is at a very low temperature $\sim 0.3$~keV -- much lower than expected for a
disc which extends down to 3 Schwarzschild radii, accreting at close to 
(or more probably somewhat above) Eddington limit. The hard component looks 
like a comptonized spectrum where the electrons have a fairly low 
temperature since the high energy
roll-over can clearly be seen.  The parameters of the comptonizing cloud are 
then fairly well constrained. It has to be optically thick, 
$\tau=5-6$, with electron temperature $\le 10$ keV. These electrons scatter 
an internal source of soft seed photons whose spectral distribution is 
broader than a blackbody, with maximum temperature $\ge 1$ keV. 
These are {\it not\/} the same as the much lower temperature 
($\sim 0.3$~keV) observed soft excess.

If the seed photons are from the accretion disc
then the comptonizing cloud is probably an optically thick corona overlying the
accretion disc. This then leads us to a picture where the inner disc is 
covered by the optically thick comptonizing layers. 
A rather large extent of the cloud is suggested by simple energetic 
arguments: since the comptonized spectrum
carries $\sim 50$ per cent of the total luminosity, its radius can be expected
to be $\sim 25\, \Rg$, if pure radial stratification of the flow is assumed.  
We see the normal disc spectrum only from larger radii, which is why the 
observed soft excess temperature is so much smaller than the $\sim 1$ keV 
expected from a $12 \MSun$ black hole accreting at Eddington rates.

The only problem with this picture is that the detected reprocessed 
spectrum is  strongly smeared, to the extent expected if it were reflecting 
from a disc which extended all the way down to $6\Rg$. However,
the constraints from relativistic smearing can be alleviated if an
additional comptonization of the reprocessed component is postulated.
Introducing purely phenomenological extension of our model,
where the reprocessed component
is additionally comptonized in a plasma of temperature $\Tr$ and 
optical depth $\taur$
(using the Green's function of Titarchuk 1994 in disc geometry), we obtain
a good fit, with $\chi^2=19.3/25\,$dof, and the additional plasma
parameters $\Tr \sim 3.7 $keV and $\taur \sim 2.5 $. The reflection amplitude
is now $f\sim 0.7$ and the inner disc radius is now unconstrained.

We note that our implementation of the additional comptonization does not
correspond to the effect recently emphasized by Ross, Fabian \& Young
(1998). They point out that the reflected X-rays below $\sim 10$ keV will be
comptonized when diffusing through strongly ionized outer layers of the
reflecting disc, an effect that is not accounted for in \relrepr model (more
accurately, this additional comptonization is included in the line profile
computations but not in the continuum). While this effect can also be important
in our case, as the reflector is indeed strongly ionized, we impose the
additional comptonization on the entire reprocessed component, thus changing 
its overall shape, rather than only below 10 keV.

\section{Spectra influenced by strong photo--electric absorption}
  
\label{sec:heavyabs}

\subsection{The initial variability}

\begin{figure}
\epsfxsize = 0.45\textwidth
\epsfbox[30 180 550 690]{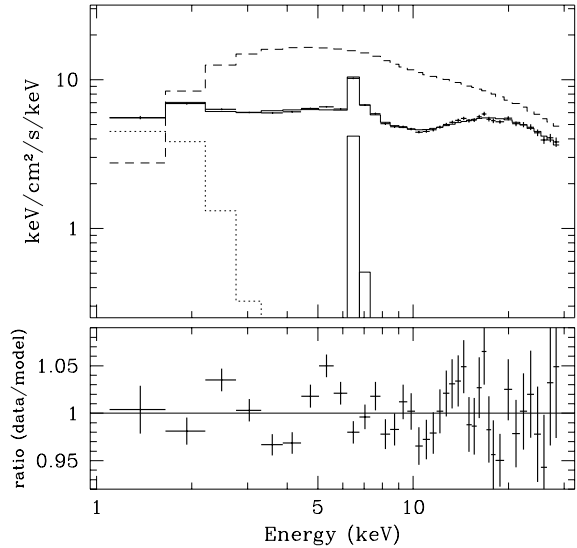}
\caption{Source spectrum (upper panel) on 30 May, 4:36:42 -- 4:36:48, 
($\Delta t=6$ sec; 329 sec after the beginning of our data), when
the flux dropped by a factor of 3 after an initial, very stable level
(cf.\ the light-curve in Figure~\ref{fig:may30_lc}). The spectrum can be 
described as strongly absorbed spectrum used for the first 200 sec
(Figure~\ref{fig:may30spec}; Section~\ref{sec:may30}). The hydrogen 
column density is 
$\NH\approx 3.5\times 10^{24}\,{\rm cm^{-2}}$, ionization parameter of the
absorber $\xi_{\rm abs}\sim 100$ and the covering fraction $\sim 0.62$.
The uppermost dashed curve shows the unabsorbed hard component.
Lower panel shows the ratio of data to best fit model. The fit is
formally unacceptable, $\chi^2=79/33\,$dof, but the residuals do not
exceed 5 per cent. They suggest that the absorption cannot be modelled
in a single--zone approximation used here.
\label{fig:s329_abs}}
\end{figure}

After the steady 200 seconds, the flux from \gs dropped dramatically. 
Is this intrinsic variability, or is it caused solely by photo--electric
absorption? 
As a first example we show a 6--seconds averaged spectrum beginning at
$t = t_0 + 329\,$s (Figure~\ref{fig:may30_lc}, \ref{fig:colcol}). 
The spectrum is plotted in Figure~\ref{fig:s329_abs}, together with the
unabsorbed spectrum discussed in Section~\ref{sec:may30}. Plainly they are
rather similar at the highest energies, but with a factor of $\sim 3-10$
deficit of counts below 15 keV in the later spectrum. This deficit is far 
too gradual a function of
energy to be caused by complete covering by neutral material, so we first 
use partial covering by heavy absorption, {\tt thabs}, on a good description 
of  the unabsorbed spectrum (model C in Section~\ref{sec:may30}:
disc blackbody and the {\tt thComp} comptonization model with consistent 
reprocessing). All the unabsorbed spectral parameters are frozen, except 
for its
overall normalization. This gives a very poor fit to the data, with
$\chi^2=2028/36$, but including an additional neutral absorber gives a 
much improved fit with $\chi^2=634/35$, where the absorption parameters 
are: complete covering with $\NH\sim 1.6\times 10^{22}$ cm$^{-2}$, and partial
covering with $\NH\sim 3\times 10^{24}$ cm$^{-2}$ of 65 per cent of the source.
However, the fit underestimates the spectrum in the iron line region and 
at energies beyond 10 keV.
Allowing the additional gaussians to be free gives $\chi^2=130/34$, while
including the amount of reflection in the fit
gives $\chi^2=77/33$, for an reflected fraction of
$\Or\sim 0.7$ and line EW of $\sim 290$ eV (compare with $\Or\sim 0.17$ and EW
$\sim 70$ eV in the unabsorbed spectrum). The complex neutral absorption then
has $\NH \sim 10^{22}$ cm$^{-2}$ for complete covering and $\NH\sim 2 \times
10^{24}$ cm$^{-2}$ covering 65 per of the source. 

The apparent changes to the reflection and line emission can be caused by our
use of {\tt thabs} to describe the very heavy absorption. At such high columns,
Compton down-scattering can distort the shape of the absorbed spectrum. We
replace the analytic approximation by our Monte Carlo transfer code to model
the heavy partial absorption. With only the line emission allowed to be free we
obtain $\chi^2=90/34$, with absorption parameters $\NH\sim 1.4\times
10^{22}$ cm$^{-2}$, and partial
covering with $\NH\sim 3\times 10^{24}$ cm$^{-2}$ of 62 per cent of the source.
For this model, the normalization of the underlying spectrum is less than 
5 per cent 
different to that of the unabsorbed spectrum except that the addition (narrow
gaussian) iron line intensity has increased by a factor of $\sim 3$. 

The fit can be improved somewhat by allowing the heavy partial absorber to be
ionized. This is the best fit spectrum shown in Figure~\ref{fig:s329_abs}. The
column density is $\NH=3.5\times 10^{24}\,{\rm cm^{-2}}$ i.e.\ $\taues\approx
2.8$, $\xi_{\rm abs}\sim 100$ and the covering fraction $\sim 0.6$.  However,
even this gives $\chi^2\sim 79/33$, i.e.\ the fit is formally unacceptable with
residuals at the 5 per cent level. The fit cannot be improved by small changes
to the parameters of the primary (comptonized) spectrum or the soft component.

Thus even though the overall spectrum is fairly well explained
by the absorption hypothesis, the presence of significant residuals points
towards complexity of the absorber. Since the data are an average over 
a long enough time for the absorber to change, a model involving a range
of column densities may be required, rather than a single--zone approximation
used here.

\subsection{All variability}

\begin{figure*}
 \epsfxsize = \textwidth 
 \epsfbox[18 144 592 718]{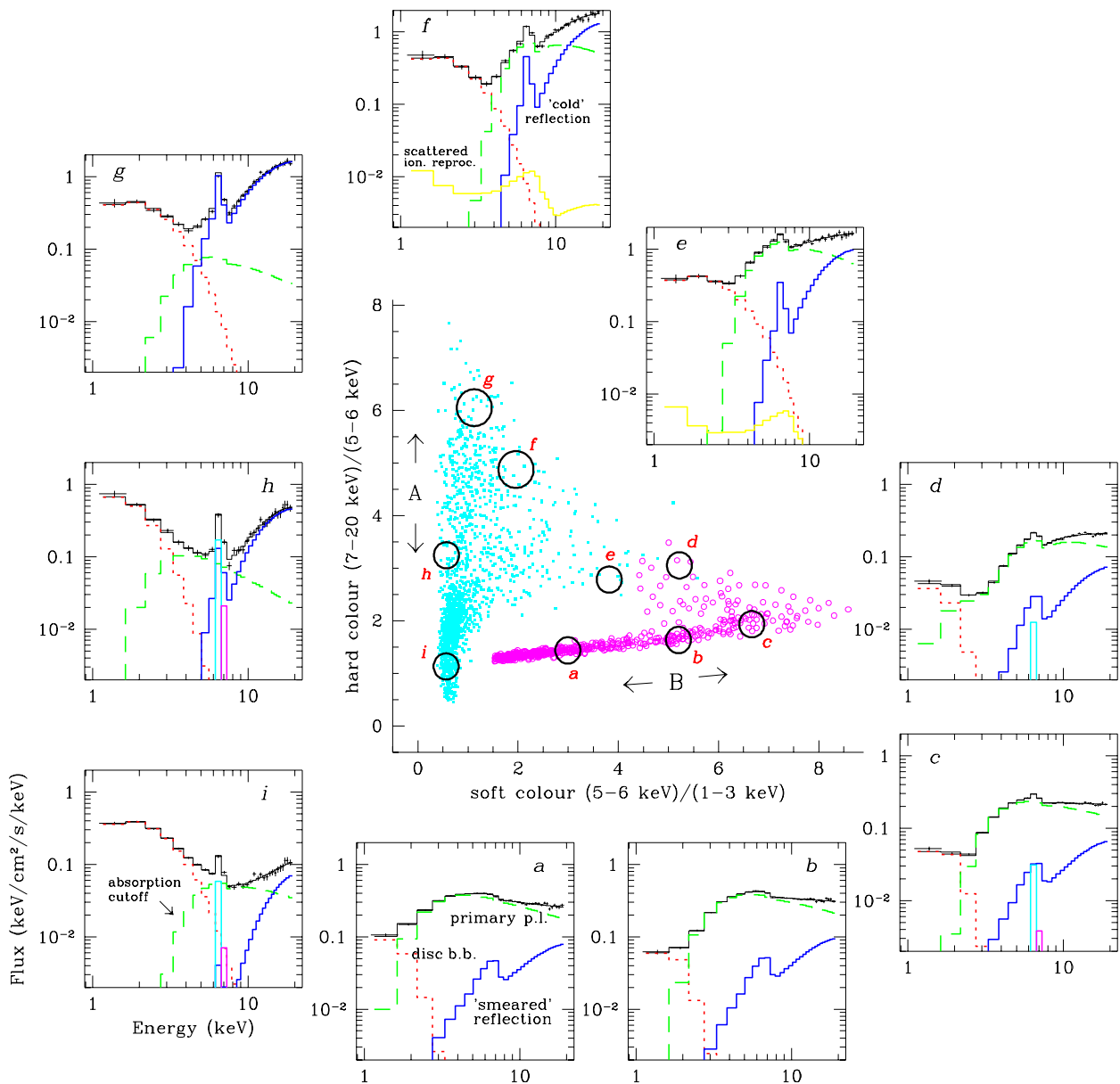}
 \caption{
The colour--colour diagram (cf.\ Figure~\ref{fig:colcol}) and examples of 
spectra influenced by heavy photo--absorption. Solid squares
are the source's positions on May 30th, 
6:30--10:00 (i.e.\ only part of the May 30th data is represented here); 
small open circles: June  3rd -- 5th. 
The labelled circles mark approximate locations where the spectra
were extracted. The spectra are assumed to consist of a soft 
component modelled as a disc blackbody, absorbed primary power law and 
its corresponding (but independently absorbed)
reprocessed component, together with gaussian lines at 6.4 keV and 7.05 keV (Fe
fluorescence), as labelled. 
Track 'A' is consistent with both the power law 
and the reprocessed component being absorbed below $\sim 4$ keV. Movement
downwards along the track corresponds to gradual disappearance of the 
reprocessed component. Horizontal movement corresponds to changes of the
normalization of the primary power law.
Track 'B' begins with unabsorbed, low state spectra (position: 1.5, 1.2)
and increasing soft colour corresponds to stronger absorption, influencing 
the primary power law and its reprocessed component.
\label{fig:heavyabs}}
\end{figure*}

We use the colour--colour diagram as a guide to the variability. 
On May 30th the
source showed dramatic spectral variability (Figure~\ref{fig:colcol}), and this
continued during the first month of the decline from outburst
(Figure~\ref{fig:heavyabs}).  The unobscured spectra are easy to spot on such
diagrams, since they fall at the leftmost end of a distinct horizontal track in
the colour--colour plots (see also Oosterbroek et al.\ 1997).  The unusual
spectrum seen at the unobscured peak outburst luminosity of \gs is {\it not\/}
 the
same as the unobscured spectra seen several days later (see Paper I). The
unobscured spectra at the end of the absorption track from June 3rd onwards are
fairly typical low/hard state spectra (see Paper I), and have inferred
luminosities of $\le 5$ per cent of $\LEdd$. Hence there must 
be some intrinsic
spectral and intensity variability as well as the photo--electric absorption. 

We will not attempt to construct a full time history of the absorber but will
limit our efforts to identifying the spectral components present when the 
source was in various locations on the colour--colour diagram.  
Because of the dramatic variability (often on 1 sec time scale), 
the time intervals over which we summed the spectra are a
compromise between requirements of a stable position on the colour--colour
diagram and photon statistics.  We generally assume that the intrinsic spectrum
consists of three components: a soft thermal component which may or may not
contribute the seed photons for Compton upscattering to make the hard power 
law, the comptonized power law and its corresponding reprocessed component. 
The absorbing material,
as well as distorting the intrinsic spectrum, may produce fluorescent line
emission, or a reprocessed component of its own. 

In the May 30th $i-1$ interval we have shown that the spectral and intensity
variability is consistent with complex, heavy absorption. The $i-2$ interval 
shows similar behaviour to the $i-1$ interval on the colour--colour plot
(Figure~\ref{fig:colcol}), with strong variability in the soft colour but 
little change in the hard colour. Thus it seems likely that most of the 
variability during this interval is also caused by a changing column of 
the heavy absorption which covers 60--70 per cent of the source, while 
the intrinsic spectrum remains more or less constant.  The variability in the 
$i-3$ interval is apparently different, starting off in a very low intensity 
state which is much harder at high energies (approximate position in the 
diagram  $[1.4,2.5]$), which then rises to $[2,3]$, and then rejoins the $i-1$ 
and $i-2$ absorption track at $\sim [3,1.5]$.  However, 
even here the
spectrum is roughly consistent with that of the peak luminosity state (in both
shape and intensity), with the heavy absorption changing from a covering
fraction of $\ge 90$ per cent at the start to $\sim 70$ per cent as seen 
in the $i-1$ and $i-2$ intervals at the end.

However, the $i-4$ spectrum is completely different. It cannot be fit by
any form
of absorption of the peak luminosity spectrum. There is a distinct hard tail,
showing that there is indeed some heavy photo--electric absorption, but the low
energy spectrum is much softer than that of the peak spectrum (as shown by it
having a smaller hardness ratio at low energies). The intrinsic spectrum has
changed! We can model this spectrum using a soft component whose temperature 
is now 
$\sim 1$ keV. The hard X--ray spectrum is then compatible with these being 
the seed
photons for Compton upscattering, forming a fairly steep power law with
$\Gamma\sim 2$, and where the rollover from the electron temperature is not
detectable. This is then partially covered by heavy photo--electric absorption.
We will discuss the nature of the intrinsic spectrum in more detail in the next
section. Here we will merely note its resemblance to the classic high state
spectrum (high temperature, steep power law component). The total bolometric
unobscured luminosity is then derived to be
$\sim 7\times 10^{-8}$ ergs cm$^{-2}$, equivalent to $ 0.07 \LEdd$.

None of the data taken after this time show a significant high energy rollover,
and all have a hard component which can be described by comptonization of the
observed soft photons. We use this interval as the starting point (data taken 
after May 30th 06:30), and plot the colour--colour diagram for all the data 
up to June 5th (see Figure~\ref{fig:heavyabs}). We have 
extracted spectra from a number of regions on the colour--colour diagram,
and modelled them to try and deconvolve the intrinsic
spectrum from the effects of absorption.

During $i-4$ and $i-5$ the source changes mainly in hard colour, making 
a vertical track (track A) in the colour--colour diagram 
(Figure~\ref{fig:heavyabs}). This track continues in $i-6$ and $i-7$, but 
here the spectral variability becomes faster
and more chaotic, and includes changes in soft colour as well. 
Good fits are obtained to spectra {\it i}--{\it h}--{\it g} along track A 
with a model in which the
soft {\tt diskbb} component has little absorption, while the hard component
(Compton scattered power law tail) is more strongly absorbed. In addition, this
hard power law illuminates cool, neutral material, giving a reflected spectrum
which is even more strongly absorbed, and whose normalisation is strongly 
enhanced with respect to the observed hard component. Motion from the bottom 
of track 'A' to the top corresponds to increasing the enhancement of this 
reflected component
from about a factor 7 at spectrum {\it i\/} to about a factor 100 at 
spectrum {\it g\/}. The soft spectrum is consistent with remaining fairly 
constant throughout this track (see Oosterbroek et al.\ 1997), 
while the hard component shows the normal broad band variability
(see below).

The total unobscured luminosity at the top of track A is then $\sim 0.5 \LEdd$
so the inferred luminosity increases by a factor $\sim 10$ from the bottom to
top of track A. Yet it seems
rather unlikely that this change is actually intrinsic, since the soft 
component shows very little change in temperature or normalization. 
It is more likely that
the source remains fairly constant at $0.5\LEdd$, but that some (variable)
amount of the source is completely obscured by optically thick material, 
and that we see only the intrinsic source spectrum via scattering.

\begin{figure*}
\epsfysize = 7 cm
\epsfbox[18 420 600 690]{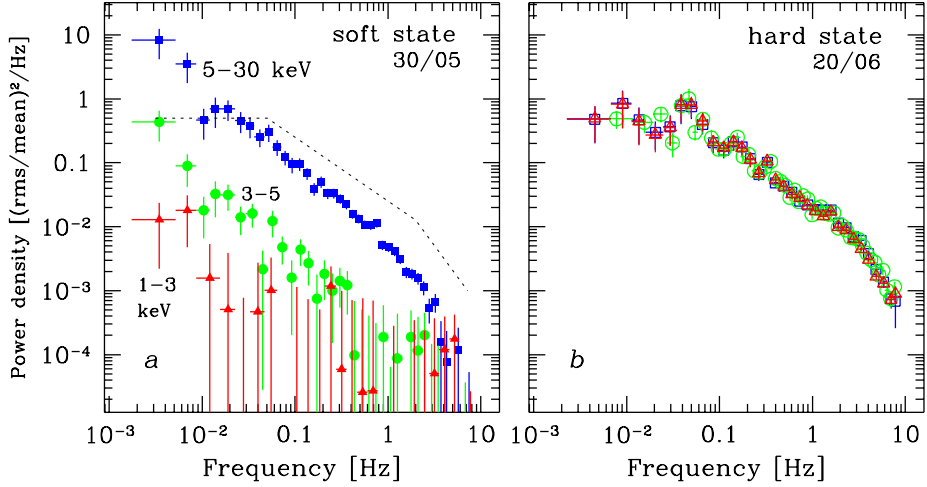}
\caption{Power spectra of X-ray emission obtained on two occasions:
panel ({\it a}\/) shows PSD on 30th May (time interval $i-5$, 07:26 to 07:48; 
see  Figure~\ref{fig:may30_lc}), panel
({\it b}\/) shows PSD on 20th June when the source was in the usual
low/hard state (Paper I). Labels in panel {\it a\/ } show energy bands in keV.
The thin dotted line represents PSD from panel {\it b\/} for direct
comparison.
 On 30th May the $1-3$ keV component did not vary on time-scales shorter than
$\sim 100$ sec, but the amplitude of variability increased with energy.
In all three spectra an enhanced variability on time-scales longer than
$\sim 100$ sec was observed. This can be attributed to variable photo-electric
absorption.
On 20th June (panel {\it b\/}) the variability was independent of energy.
 \label{fig:may30_psd}}
\end{figure*}

The scattering origin for the observed soft component would be in accord with
its  small normalization, which is only 
$\sim 1$ per cent of that expected, if the component indeed is a disc emission 
from $6\,\Rg$ onwards. We note here the strong constraints on
the scattering scenario for the hard component coming from its observed broad
band variability. The high amplitude of the PSD of the hard component 
($E>5\,$keV) during the $i-5$ interval (Figure~\ref{fig:may30_psd}) means
that the scatterer could not have been located farther than $\sim 0.1$
lsec $\sim 10^3\,\Rg$ (see also next Section).

One geometrical scenario in which to understand these results is that there 
is a thick disc which blocks our direct view of the source, so that we see 
only the scattered fraction of the intrinsic radiation. The obscuring wall is
axi--symmetric, so there is a strong reflected component of the intrinsic
(rather than scattered) emission from the opposite wall. As the disc subsides
then we see progressively more of the reflector, so the strength of the hard
component increases, giving the variability seen as track A.
Eventually the optically thick material goes below our line of
sight, allowing the intrinsic hard spectrum to be seen. Differing amounts
of fairly heavy absorption on this intrinsic hard spectrum then cause the 
variability 
in soft and hard colour defined by spectra {\it g}--{\it f}--{\it e}, but the 
intrinsic spectrum is
consistent with remaining approximately the same in both shape and intensity.

Thus despite the dramatic spectral and intensity variability, the data on 
May 30th
are compatible with very little intrinsic variability. At the start of the
observation the source was accreting at (super) Eddington luminosities, with
a spectrum unlike any of the known (sub Eddington) spectral states. It then
declined by only a factor of $\sim 2$, and made a state transition to the high
(or very high) state, and remained at this level for the rest of the 
observation.  All the rest of the variability seen is connected to the 
absorption. 

After the May 30th observation the next data were not taken until 48 hours
later, mid-morning of June 1st. These were taken with satellite pointing 
position
which was offset by $\sim 0.8^\circ$ from the source, so cannot be used for
detailed spectral analysis. However, on the colour--colour plot they span the
region around spectrum {\it d}. The next good data are taken 12 hours later, 
on the early morning of June 2nd, where the source spectrum again looks very 
similar to that of spectrum {\it d}.  At first sight this looks like a 
continuation of the {\it g}--{\it f}--{\it e}
track, but fitting of the data shows distinct differences. Firstly the overall
intensity is rather lower, with the hard component being a factor 10 weaker
than in spectrum {\it e}. The soft emission is also at much lower 
temperatures. These, together with the good data taken on June 3-5th form a 
distinct track in Figure~\ref{fig:heavyabs} along spectra 
{\it d}--{\it c}--{\it b}--{\it a} (track B)
which is {\it not\/} simply connected to track A. Again there must have been 
some
intrinsic spectral change, since the unabsorbed spectra at the end of this
track show a classic low state spectral form (Paper I). 

Track B can be reproduced by cold partial covering of the Jun 3rd unabsorbed 
low state spectrum, fixing all the parameters to be the same as in the 
unabsorbed data (see Paper I).  In this model the overall normalization of 
the June 3rd
spectrum (which gives a total bolometric flux of $\sim 4\times 10^{-8}$ ergs
cm$^{-2}$ s$^{-1}$ or $0.04\LEdd$)
changes by less than 25 per cent while the partial covering absorption
parameters change from $3 \rightarrow 7 \rightarrow 12 \times 10^{22}$ 
cm$^{-2}$ covering $\sim 92$ per cent of the source going from spectra 
{\it a}$\rightarrow${\it b}$\rightarrow${\it c}. 
This confirms the suggestion by Oosterbroek et al.\ (1997) 
that the track corresponds roughly to increasing $\NH$. We have however used
here a much more realistic intrinsic spectrum -- the
unabsorbed spectrum of the June 3rd data set (Paper I).  

The turning point of track 'B' corresponds to the absorption column of $\approx
1.5\times 10^{23}\,{\rm cm^{-2}}$. Further increasing $\NH$ leads to a
decrease of the soft colour since it decreases the counts in the 3--5 keV band,
whilst the counts in the 1--3 keV band -- determined by the soft component --
are constant. Thus spectrum {\it d\/} can be fit into the above pattern, with 
a column of $\sim 31\times 10^{22}$ covering 92 per cent of the source. 

\section{Has the source ever made a transition to a soft state?}

\label{sec:soft}

As we point out in Section~\ref{sec:heavyabs} some of the spectra during 
May 30th observation show a thermal component of temperature $\sim 0.8\,$keV
(for {\tt diskbb} model) and power law tails with $\Gamma \sim 2$, 
typical for high/soft states of GBH.
To further characterize the source in those time periods we examined more
closely some of the energy spectra as well as the source variability.

\subsection{Timing and variability}

\label{sec:soft_timing}

Power spectra of \gs were already examined by
Oosterbroek et al.\ (1997). Their Fig.~8 (panel 1A) shows PSD on May 30
exhibiting strongly reduced power above $\sim 1$ sec, similarly to
what is usually observed in high state (van der Klis 1995).
In Figure~\ref{fig:may30_psd}a we plot PSD for data obtained between
07:26 and 07:48 ($i-5$; see Figure~\ref{fig:may30_lc}), 
separately for three energy bands: $1-3\,$keV, $3-5$ keV 
and $5-30\,$keV. The $1-3$ keV band variability is consistent with pure 
Poisson noise
on time-scales $\delta t \la 100$ sec whilst the harder X-rays are highly
variable with the amplitude of variability increasing with energy up to 
$\approx 5$ keV.
For comparison Figure~\ref{fig:may30_psd}b shows PSD for data taken
on 20th June (cf.\ Miyamoto et al.\ 1992), when the source was in the usual 
low/hard state (Paper I). 
Here the amplitude of variability does not depend on energy.

This analysis confirms that the energy spectrum consists of two components,
a constant, soft component, and a strongly variable hard component, in
accord with spectral decomposition shown in Figure~\ref{fig:soft_spec}.

A characteristic feature of the PSD is the increase of power for
$\delta t>100$ sec. Since even the soft component varies on that time-scale,
the increase is most likely due to slowly variable absorption.

The PSD of the hard component (5 -- 20 keV) on May 30th is rather steeper 
than that on 20 June for $\delta t \la 10$ sec (frequency $f \ga 0.1$ Hz). 
The loss of power could be due to reflection/scattering 
in an extended medium whose presence is suggested both by the energy spectra
analyzed in the next section and by the previously discussed increase 
of PSD on long time-scales. Alternatively, it could be an intrinsic feature
of the primary emission. If so, however, then the steepening of PSD in
putative soft state would be opposite to what is shown by e.g.\ Cyg X-1
(van der Klis 1995 and Cui et al.\ 1997).

\subsection{Energy spectra}

\begin{figure*}
\epsfysize = 7 cm
\epsfbox[18 440 590 690]{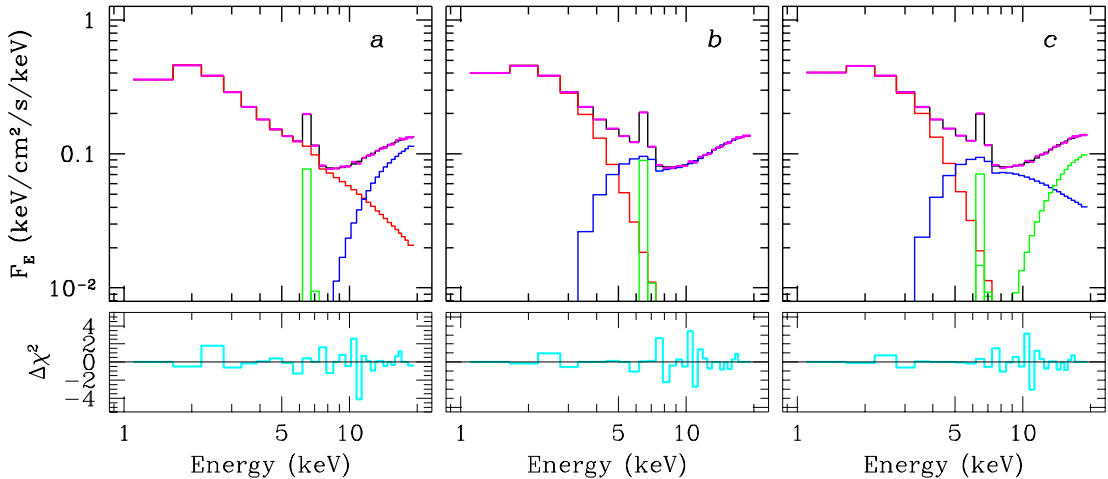}
\caption{Examples of spectral modelling of the $i-4$ spectrum, when the source
was heavily absorbed. Model {\it a\/} is
purely phenomenological, comprising two power law spectra absorbed by two
(partial) absorbers and an overall absorption. Model {\it b\/} consists of
an unabsorbed disc blackbody with $k T\approx 1$ keV, and a 
power law ($\Gamma\approx 2$) with two absorbers 
($N_{\rm H,1}\approx 24$, $f_1=1$; $N_{\rm H,2}\approx 300$, 
$f_2=0.86$). Model {\it c\/} has similar disc blackbody,
a power law with $\Gamma=2.3$ absorbed by $\NH=27$ and $f=1$, and an 
(enhanced) reprocessed component absorbed by $\NH=180$ and $f=1$ (all values 
of $\NH$ are given in units of $10^{22}\,{\rm cm^{-2}}$). Additional narrow
gaussian lines at 6.4 and 7.05 keV are included to account for fluorescence
in the absorbers. Values of $\chi^2$ are, respectively: $21.2/22\,$dof,
$20.6/23\,$dof and $17.4/22\,$dof. Only models 
{\it b\/} and {\it c\/} are compatible with PSD shown in 
Figure~\ref{fig:may30_psd}a, i.e.\ the variability increasing with energy
up to $\sim 5\,$keV. Parameters of those models are typical for high/soft
state of SXT.
\label{fig:soft_spec}}
\end{figure*}

We have extracted two spectra of \gs from  time intervals: $i-4$ at
6:43:45 -- 6:47:20 and $i-5$ at 7:26:45 -- 7:49:00
(see light-curve in Figure~\ref{fig:may30_lc}, and position on the 
colour--colour diagram, Figure~\ref{fig:colcol}). 
First, we tested the hypothesis that the spectra can be described
assuming (possibly non-uniform) absorption acting on a typical hard state
spectrum. To this end we assumed the June 3rd (see Table~2 in Paper I) 
spectrum and allowed the three
components (disc blackbody, power law and reflection) to be absorbed. This
model fails in both cases even though we allowed the three absorbers to be
different. The best fits have $\chi^2=1998/22\,\dof$ and 
$\chi^2=571/22\,\dof$. 
Adding to the model two narrow gaussians at 6.4 and 7.05 keV (to account for 
iron fluorescence from the absorber) improves the fits significantly but
they are still unacceptable: $\chi^2=1220/21\,\dof$ and 
$\chi^2=56.3/21\,\dof$. 


\begin{table*}
 \caption{Model fitting of the soft state spectra.
\label{tab:soft_spec}}
   \begin{tabular}{lcccccccr}
 data & $k T_{\rm soft}$ (keV) & 
        photon index & $N_{\rm H,1}\,{\rm cm^{-2}}$ &
        $\Or$ & $\xi$ & 
	$N_{\rm H,2}$ &
        EW (eV) & $\chi^2/$dof \\
 \hline
$i-4$ & $1.07^{+0.04 }_{-0.11}$ & 
        $2.24\pm 0.16$ & $(25^{+5}_{-8})\times 10^{22}$ &
        $7.4^{+2.2}_{-1.1}$ & $100^{+400}_{-100}$  & 
        $(180^{+30}_{-20})\times 10^{22}$ &
	$430^{+70}_{-110}$ & 17.4/21 \\
$i-5$ &  $0.98^{+0.03 }_{-0.06}$ &
        $1.90^{+0.13}_{-0.11} $ & $(24\pm 5)\times 10^{22}$ &
        $6.3^{+1.0}_{-0.8}$ & $25^{+40}_{-24.8}$ &
        $(130^{+27}_{-20})\times 10^{22}$ &
        $375^{+75}_{-60}$ & 15.8/21 \\
 \hline
   \end{tabular}

\medskip

Model: abs*( disc blackbody + abs1*(power law) + abs2*(\relrepr) + 
gaussian + gaussian) 

\end{table*}

The simplest phenomenological description of the two spectra consists of 
two power law components with different absorbing columns, $\NH$, and covering 
fractions, $f$, with the two narrow gaussian lines 
and an overall absorption.
One power law is  rather soft, $\Gamma\sim 3$ and it is absorbed by a 
column of $\NH\sim (6-8)\times 10^{23}\,{\rm cm^{-2}}$ and $f = 0.65-0.9$. 
The other is harder, $\Gamma\sim 1.2$ while $\NH\sim 2\times 10^{24}\,
{\rm cm^{-2}}$ and $f=1$. 

In an attempt to construct a more  realistic model we assumed an
unabsorbed disc blackbody as the soft component and an absorbed 
power law, but such a model fails to describe the data. Better fits are 
obtained when a second, partial absorption is applied to the power law:
$\chi^2=20.5/22\,$dof for $i-4$ and $\chi^2 = 27.8/22\,$dof for $i-5$,
with the second absorbing column $\sim 3\times 10^{24}\,{\rm cm^{-2}}$.
The fit to $i-5$ data can be further improved if the second absorber acts
on a different power law component. One realistic candidate for the 
second power law is the reprocessed component whose presence
is also in line with our previous results. Replacing the power law with 
the (non-smeared) reprocessed component we obtain
good fits with $\chi^2=17.4/21\,\dof$ for $i-4$ and $\chi^2=15.8/21\,$dof 
for $i-5$. 

The model parameters are well constrained: the temperature of 
the soft component is fairly high, $T_0 \sim 1\,$keV, 
whilst the power law index is $\sim 2$.
Both values are indeed as expected in typical high/soft state spectra
of GBH.
The spectra are plotted in Figure~\ref{fig:soft_spec} and fit results shown
in Table~\ref{tab:soft_spec}.

\subsection{Discussion}

Based solely on properties of the $i-4$ and $i-5$ spectra it is impossible
to determine whether the constancy of the soft component is its intrinsic 
property or it results from the component (or a fraction of it) being 
scattered off some extended scatterer towards the observer. 
The lack of variability is intrinsic if the hard component is scattered 
as well -- as the overall evolution seems to suggest -- since the scattering
preserved the variability characteristics of the hard component. 
An intrinsically weak variability then supports the spectral identification 
of the  soft component as the thermal disc emission since this is known
to be weakly variable (van der Klis 1995).

The intrinsic luminosity of the soft component is $\approx 9\times 10^{38}
\,{\rm erg\,cm^{-2}\,s^{-1}} \approx 0.6\,\LEdd$ ($d=3.5\,$kpc), assuming 
that it indeed is a disc emission from $6\,\Rg$ onwards. Effective temperature
expected at the inner disc edge is then $\approx 1.1\,$keV (Frank, King \& 
Raine 1992),  in surprisingly good agreement with the best fit value
(fitting the {\tt diskspec} model with the colour temperature correction
1.5).

The energy spectrum of \gs during the $i-4 \rightarrow i-7$ time intervals
is thus compatible with typical high/soft state spectra of GBH. However,
the strong, broad band noise variability of the hard component is not. 
Examining more closely the {\it Ginga\/} data of Nova Muscae 1991 (see
Section~\ref{sec:compar}) we find that in the typical high/soft state the hard 
component 
($E>5\,$keV) shows almost no variability, similarly to the soft component
(note that in the {\it Ginga\/} data, counts are dominated by channels
at 3--5 keV, so the usually plotted -- summed over all energies -- PSDs 
are dominated by the soft component). Perhaps a weak intrinsic variability 
is enhanced by absorption, as suggested by the fact that the 
 PSD amplitude of the soft component was larger later, in $i-7$ than in
$i-4$ and $i-5$.

\section{Discussion and Conclusions}

\subsection{Intrinsic X--ray Spectral Evolution}

Much of the evolution of the X--ray spectrum is hidden beneath a veil of 
complex, heavy absorption, and it is not generally possible to uniquely 
recover the intrinsic
spectrum. Most of the dramatic spectral {\it and intensity\/} variability 
seen on
May 30 is connected with the evolution of the complex absorption rather than
with the X--ray source itself. The apparent saturation of the X--ray luminosity
at $\sim 10^5$ counts sec$^{-1}$ (see e.g.\ Figure~\ref{fig:may30_lc}) 
is {\it not\/} due to the source
dramatically flaring and then hitting the Eddington limit, but instead can be 
explained by the source staying fairly constant while our direct line of sight
 to it is covered (and uncovered) by very optically thick material. 

The absorption variability is such that there are occasional glimpses of the
unobscured source. One such time is at the start of the observation on May
30. Here the observed luminosity is at least 0.6 of the Eddington limit 
(integrating the
derived spectrum to get the bolometric luminosity gives the model dependent
number of $\ge 1.6\times\LEdd$), and the spectral shape is not at all like
that seen from other transient systems (at any luminosity!).  At these high
accretion rates we might expect to see a strong soft component from the inner
accretion disc at around $\sim 1$ keV, yet the observed strong soft emission is
at temperatures around $0.2$ keV, much lower than expected. The hard X--ray
spectrum is not a power law. Instead it has a distinct roll-over, indicating
relatively low electron temperatures, $\sim 10$ keV, in the hard X--ray 
emitting
plasma. The curvature of this spectrum is rather difficult to match unless the
seed photons for this comptonized spectrum are at temperatures of $\sim 1$ keV.
Thus it seems most likely that the missing accretion disc photons are hidden
under the optically thick comptonizing cloud which produces the hard X--ray
spectrum.  Slim disc models (including the advection of trapped radiation) of
super-Eddington rates may point towards the formation of such a hot central
region (Beloborodov 1998). These extreme luminosities are probably accompanied 
by
a strong wind driven from the disc, which is perhaps the origin of some of the 
dramatic absorption variability.

The source then makes a transition to the
standard very high or high state spectrum, with the expected strong soft
component at $\sim 1$ keV. This transition probably takes place between the
$i-3$ and $i-4$ time intervals (see Figure~\ref{fig:may30_lc}), 
perhaps connected to the 
source luminosity decreasing from $\sim 1.5\times \LEdd $
to slightly below $\LEdd$. All the May 30th spectra are consistent
with an intrinsic bolometric luminosity close to $\LEdd$, although this is
highly model dependent due to the obscuration of the source. 

The source was not observed on May 31, and no good data exist for the 
observation on June 1st. This is somewhat unfortunate, since on June 2nd
the data are consistent with the source having an intrinsic luminosity of 
$\sim 0.04-0.05 \,\LEdd$, 
and show the standard low/hard spectrum. Somewhere in the two missing
days the intrinsic source luminosity decreased by a factor of 10--20! 
The source then decreases by about a factor of 2 from June 2nd to July 6th,
consistent with a standard e--folding decay timescale of 30--40 days.  

In summary, all of the oddities of \gs may perhaps be explained by the 
source accreting at super-Eddington rates. This would give the unusual
spectrum seen at the start
of May 30th, and power a strong outflowing wind which caused dramatic 
absorption
variability. As the source declined below Eddington luminosity it showed 
the standard 
high state spectrum. It may have further declined steadily (although rapidly)
through the high state spectrum to the low/hard state on May 31st--June 1st, or
there may have been a more dramatic event, perhaps linked to the observed
transient radio emission, in which there was complete disruption of the inner
disc. Perhaps this signaled a huge ejection of the accreting matter, so that
the accretion rate onto the central object was much reduced, and adding to the
complex absorption.

\subsection{Disc Evolution}

\label{sec:discuss}

The most promising explanation of SXT outbursts seems to be the
classical disc instability model, modified to include the effects of
X-ray irradiation (e.g.\ Cannizzo 1993, van Paradijs 1996, King \&
Ritter 1998). The quiescent disc builds up from matter accreted from
the companion star until the outburst is triggered. The disc switches
into the hot, ionized state, and has the familiar Shakura--Sunyaev
structure. The outer disc temperature eventually drops below the H
ionization temperature, and a cooling wave propagates inwards,
switching the disc back into quiescence. Without X--ray irradiation
the outburst lightcurves are the classic drawf nova lightcurves, with
a linear decline after outburst. However, if X--ray irradiation is
strong enough to keep the outer disc ionised then the cooling wave is
supressed. Most of the disc mass can then accreted before the cooling
wave can form, giving an exponential decay (King \& Ritter 1998, King
1998). The prevalence of exponential decays in SXT X--ray lightcurves
point to the importance of irradiation in these systems 
(Shahbaz, Charles \& King 1998).

\gs shows a fast rise, followed by an exponential X--ray decay.
The fast rise suggests that the outburst is triggered towards the
outer edge of the disc (Smak 1984, Cannizzo 1998), so that all of the disc
takes part in the initial heating wave. The exponential decay means
that we might expect that irradiation (probably indirect via
scattering in a corona: Dubus et al.\ 1999) is important so that most of the
disc stays in the outburst state, and so is accreted.  Integrating the
observed X-ray lightcurve gives an estimate of the accreted mass
$\DMX \sim 6\times
10^{25}\,$g, assuming a radiative efficiency of 10 per cent. This
matches very well with estimates of the mass transferred from the
companion star during the inter-outburst time interval $\DMC \approx
2\times 10^{26} $g, adopting $\Delta t \approx 32\,$years; (Chen et
al.\ 1997) and the mass transfer rate from the companion star $\dot
M_{\rm c}\approx 10^{17}\,{\rm g\,s^{-1}}$ (King, Kolb \& Burdieri
1996). 

However, this disc mass is orders of magnitude smaller than the disc
mass that would be built up under the assumption that the outburst is
triggered by the surface density of the quiescent disc reaching its
critical value everywhere. The orbital period of \gs is 6.5 days, so
the disc can
extend out to radii of $\Rout \simeq 1.36\Rcirc \approx 0.7 \RT
\approx 1.2\times 10^{12}$ cm (Shahbaz et al.\ 1998). This
predicts a huge disc mass of $\Mdisc \sim \rho \Rout^3 /3 \sim 2 \times
10^{28}$ g ($\rho\sim 3\times 10^{-8}\,{\rm g\,cm^{-3}}$; King \&
Ritter 1998).  Plainly this shows that the outburst is triggered long
before this maximum disc mass is built up. 
A similar result is found for the other well studied large disc system, 
GRO J1655-40 (Shahbaz et al.\ 1998). 
This is in marked contrast the to short period
(small disc) SXT's, where the disc mass derived from the maximum
quiescenct disc calculations is (roughly) equal to the mass inferred
from the X--ray luminosity (Shahbaz et al.\ 1998), and to the integrated
mass transfer rate from the companion star (Menou et al.\ 1999).
Perhaps a clue to resolving the problem is that the region where the
disc solution is unstable in \gs is quite distant from $\Rout$.
Solving the vertical disc structure equations (using the code
recently described in R\'{o}\.{z}a\'{n}ska et al.\ 1999), we find that
the solution is unstable where $\Teff=(5 - 7) \times 10^3\,$K
(see also e.g.\ Hameury et al.\ 1998). This gives 
$\Rst = (5 - 8) \times 10^{10}\,$cm, 
i.e.\ the ring is centered at $\approx 0.05\,\Rout$. In Nova Muscae 1991
($P_{\rm orb}=10.5\,$h) $\Rst$ is rather closer to $\Rout$:
$\Rst = (2 - 3) \times 10^{10}\,{\rm cm} = (0.2 - 0.3) \Rout$.
Perhaps the disc beyond a few $\Rst$ never builds up to
full quiescence, but instead stays on the steady state, cool branch.
Whatever the reason, it seems that there are serious deficiencies
in our understanding of the structure of large discs.

\section*{Acknowledgments}

We thank Andrei Beloborodov for helpful discussions on super-Eddington 
accretion, and Bo\.{z}ena Czerny, James Murray and John Cannizzo on 
accretion disc instabilities.
CD acknowledges support from a PPARC Advanced Fellowship.
Work of PTZ was partly supported by 
grant no.\ 2P03D00410 of the Polish State Committee for 
Scientific Research.

\appendix
\section{Differences between Monte Carlo and analytic solutions of the 
comptonization problem and their influence of modelling the reprocessed
spectra}

\begin{figure*}
 \epsfysize = 7 cm 
 \epsfbox[30 370 500 700]{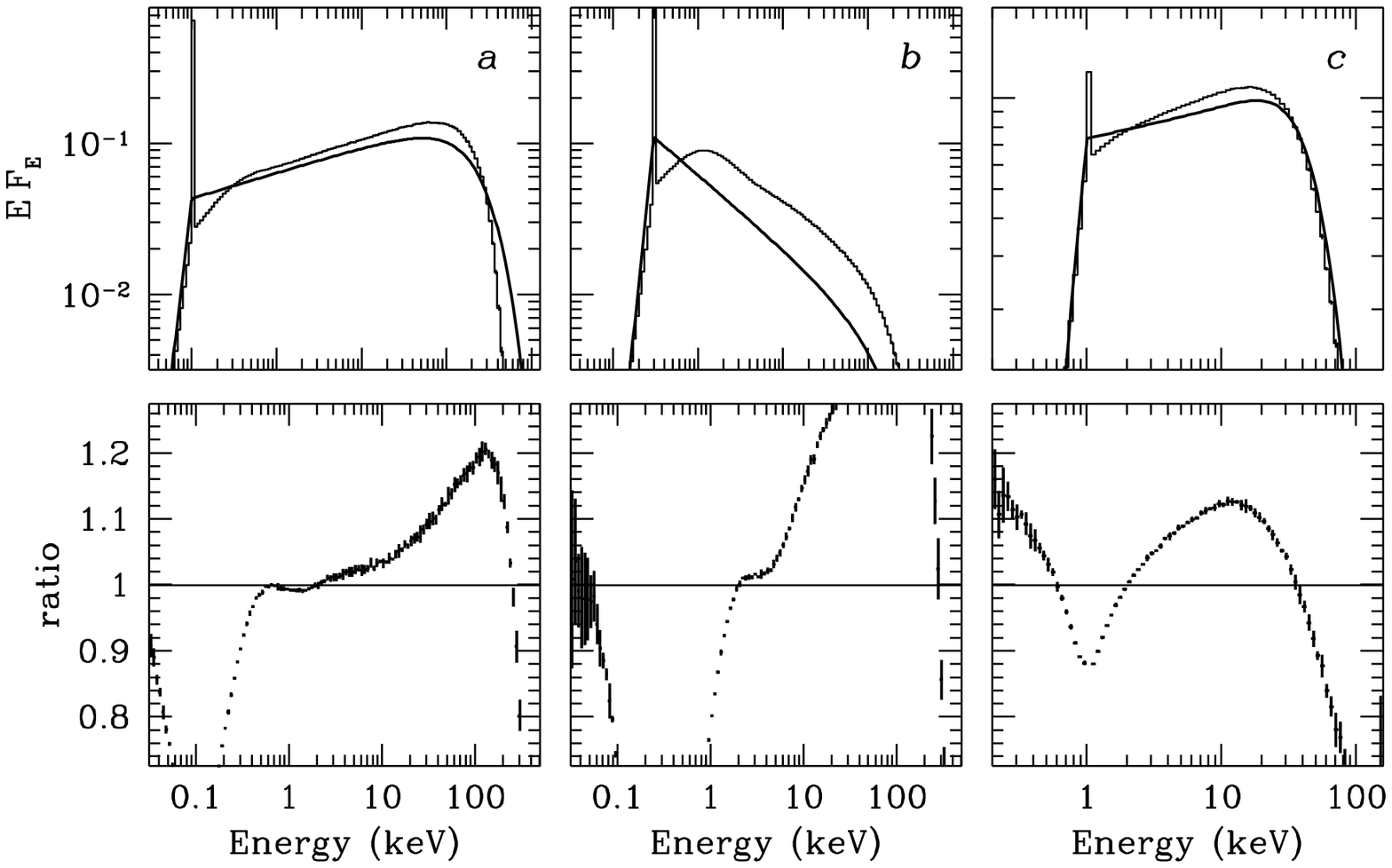}
 \caption{
Comparison of the best analytical (Titarchuk 1994) and Monte Carlo solutions 
of comptonization
Green's functions. Panel (a): $k T_0=0.1\,$keV, $\kT=100\,$keV, $\tau=1$;
(b): $k T_0 = 0.3\,$keV, $\kT=100\,$keV, $\tau=0.4$;
(c): $k T_0=1\,$keV, $\kT=10\,$keV, $\tau=6$. Upper panels show
the Green's functions, lower ones their ratio (normalized to 1 at 2 keV). 
The comptonizing
cloud is assumed to be spherical, and seed photons spatial distribution follows
$\propto \sin(\tau)/\tau$. Errors on Monte Carlo spectra
are standard deviations of 6 averaged spectra, each using $2\times 10^6$
({\it a\/}), $4\times 10^6$ ({\it b\/}) and $10^6$ ({\it c\/}) photons.
\label{fig:green}}
\end{figure*}

\begin{figure*}
 \epsfysize = 7 cm 
 \epsfbox[30 370 500 700]{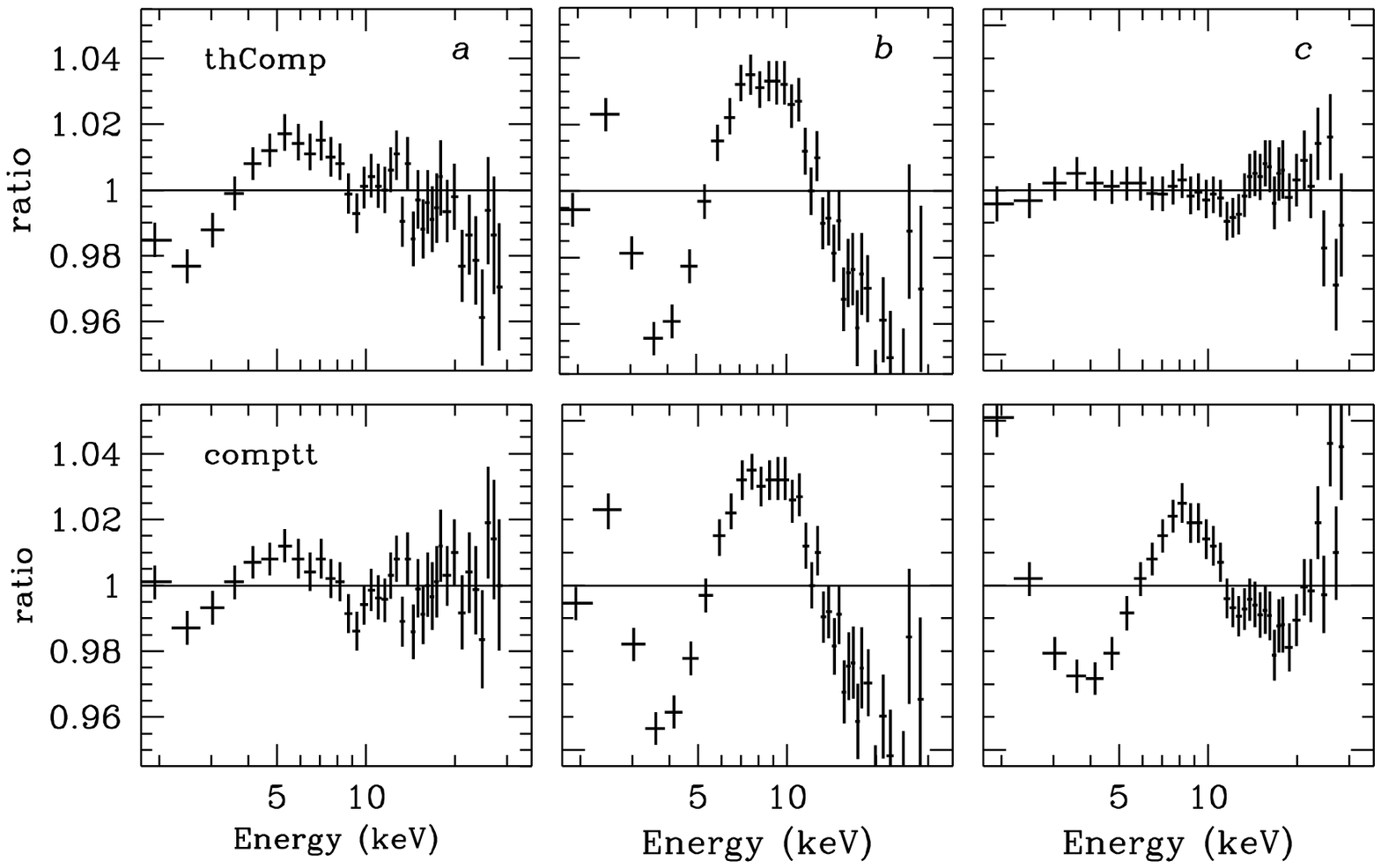}
 \caption{
Results of fitting 'fake' data created from Monte Carlo comptonization
spectra by two analytical comptonization models: {\tt thComp} and {\tt comptt}.
Parameters of the Monte Carlo spectra are the same as in Figure~\ref{fig:green}
except for the seed photons spectrum which is a blackbody here. As can be seen,
the Monte Carlo spectra can usually be quite well reproduced by adjusting 
parameters in the analytical models.
\label{fig:MCvsA}}
\end{figure*}

Analytical approximation to comptonization problem are usually not accurate
at energies close to the seed photons or plasma temperatures 
(e.g.\ Skibo et al.\ 1995).
We have therefore investigated whether the differences between ``accurate''
spectra, computed using a Monte Carlo code, and analytical approximations
(which we normally use in data fitting) may influence our conclusions
regarding the reprocessed component, in particular the inferred ionization
and the level of relativistic smearing.

Our Monte Carlo comptonization code is written following closely the 
description given by Pozdnyakov, Sobol \& Sunyaev (1983) and 
G\'{o}recki \& Wilczewski (1984). One modification we have made is to replace
the analytical approximations to 
$F(x) \equiv (2\pi r_0^2)^{-1}\, \int_0^x\, y \sigma(y)\, dy$ (where 
$\sigma(x)$ is the scattering cross-section as function of photon energy) 
and its inverse function (see Eq.\ A8 and A13--A15 in 
G\'{o}recki \& Wilczewski 1984) by values computed numerically and then
interpolated.

To demonstrate the (in)accuracy of analytical solutions of the comptonization
problem we first show in Figure~\ref{fig:green} the Green's functions  
for three cases: {\it a\/} $\kT = 100\,$keV, $\taues=1$, $k T_0=0.1\,$keV 
(typical
low/hard state spectra of GBH), {\it b\/} $\kT = 100\,$keV, $\taues=0.4$, 
$k T_0=0.3\,$keV (high/soft state of Nova Muscae 1991; \.{Z}ycki, Done
\& Smith 1998) and {\it c\/} 
$\kT = 10\,$keV, $\taues = 6$, $k T_0=1\,$keV (peak spectrum of GS~2023+338;
Section~\ref{sec:may30}). In all cases the comptonizing cloud is a uniform
sphere, with sources' distribution given by $\propto \sin(\tau)/\tau$ 
(Sunyaev \& Titarchuk 1980). The analytical model is that of Titarchuk (1994)
which we found best agrees with the Monte Carlo spectra for given 
$\kT$ and $\tau$ in the broad range of parameters we are considering.

The most certain way of testing possible influences would obviously 
be to fit the data by a Monte Carlo model and compare the results with those
obtained with analytical approximations. This being impossible in practice, 
we adopted following procedure: we created fake 'data' from comptonization 
spectra computed using our Monte Carlo code. The {\it Ginga\/} response matrix
(for summed top and mid layers; Turner et al.\ 1989)
was used to create the data and their statistical quality is similar to 
our real data. We then attempted to fit the 
analytical models to the data in the 2-30 keV range. 
We assume the same parameters as for 
the Green's functions except for the seed photons spectrum for which we now
assume a blackbody. 

Figure~\ref{fig:MCvsA} shows results of fitting two analytical models:
{\tt thComp} and {\tt comptt} to the fake data. In the cases {\it a\/}
and {\it b\/} we fixed the plasma temperature in the models to the value
used in Monte Carlo simulations since using data only up to 30 keV does
not allow $\kT$ to be constrained. The best fit optical depth is then
$\approx 1.2$ in case {\it a\/} and 0.5-0.7 in case {\it b\/}.
As can be seen, residuals up to 4 per cent can still be present. High
temperature, optically thin spectra are most difficult to reproduce as 
the analytical models usually assume a diffusion approximation for photon
escape. Fitting the data corresponding to low/hard state gives much better
results, with residuals not exceeding 2 per cent. We note remarkably good
fit of the {\tt thComp} model to Monte Carlo spectrum characteristic to 
GS~20223+338 at its peak (panel {\it c\/}), with the model parameters
within $\sim 10$ per cent of the original values. The reason for the bad fit
of the {\tt comptt} model in this case seems to be that the seed photons
spectrum is taken as the Wien spectrum rather than a blackbody used 
originally.

The residuals observed in the above fits may indeed have some influence 
on results of modelling 
the reprocessing. To asses the influence, we attempted to improve the 
fits by adding a reprocessed component to {\tt thComp} model. Results
of fitting real data by models containing reprocessing could be affected,
if indeed such a composite spectrum was able to mimic the complexity
of 'real' comptonization. This does not however seem to be the case.
For low/hard state spectra  we found that the reprocessed component 
(with $\xi$ fixed at 0, as for typical real data) cannot improve the fit,
i.e.\ the difference between Monte Carlo and analytical comptonization
cannot be 'filled in' by a reprocessed component. For high/soft state
we indeed see that an ionized and smeared reprocessing, with amplitude
$\Or\approx 0.05$ can reduce some of the residuals. We thus added such
a component to our basic model of reprocessing and we repeated fits
to high state spectrum of Nova Muscae 1991, obtained on 18th May 
(\.{Z}ycki et al.\ 1998). The resulting $\chi^2$ contours in the
$\Or$--$\Rin$ plane is shifted by $\sim 0.06$ towards smaller values but its
overall shape remains unchanged. We thus conclude that the small differences
between Monte Carlo and analytical models of comptonization are not able
to change conclusions regarding properties of the X--ray reprocessed
component in our spectra.

We finally note that the above discussion is relevant only if a 
single value of electron temperature is assumed. In more realistic models, 
where  accretion flow dynamics as well as radiative transfer are considered,
a range of temperatures can be expected. This will inevitably change the
details of the spectrum and the above discussion.

{}


\begin{thebibliography}{}

 \bibitem[]{}
   Arnaud K. A. 1996, in Astronomical Data Analysis Software and 
   Systems V, eds. Jacoby G. and Barnes J., ASP Conf. Series volume 101, 
   p. 17
 \bibitem[]{}
   Beloborodov A. M. 1998, MNRAS, 297, 739
 \bibitem[]{}
   Cannizzo J. K. 1993, in Accretion Disks in Compact Stellar Systems, ed. 
                 J. C. Wheeler (Singapore: World Scientific), p. 6
 \bibitem[]{}
   Cannizzo J. K. 1998, ApJ, 494, 366
 \bibitem[]{}
  Casares J., Charles P. A., Jones D. H. P., Rutten R. G., Callanan P. J.
   1991, MNRAS, 250, 712
 \bibitem[]{}
   Chen W., Shrader C. R., Livio M. 1997, ApJ, 491, 312
 \bibitem[]{}
   Cui W., Zhang S. N., Focke W., Swank J. H. 1997, ApJ, 484, 383
 \bibitem[]{}
   Done, C., Mulchaey, J. S., Mushotzky, R. F., Arnaud, K. A.
   1992, ApJ, 395, 275
 \bibitem[]{}
   Dubus G., Lasota J.-P., Hameury J.-M., Charles P. 1999, MNRAS, 303, 139
 \bibitem []{}
   Ebisawa K., 1991, PhD thesis, Univ.\ of Tokyo
 \bibitem[]{}
   Fabian A. C., Rees M. J., Stella L., White, N. E. 1989, MNRAS, 238, 729
 \bibitem[]{}
   Frank J., King A. R., Raine D. 1992, Accretion Power in Astrophysics 
    (Cambridge:Cambridge Univ. Pres)
 \bibitem[]{}
   Gierli\'{n}ski M.,  Zdziarski A. A., Poutanen J., Coppi P., Ebisawa K.,
   Johnson W. N. 1998, MNRAS, submitted
 \bibitem[]{}
   G\'{o}recki A., Wilczewski W.  1984, Acta Astron., 34, 1
 \bibitem[]{}
   Hameury J.-M., Menou K., Dubus G., Lasota J.-P., Hur\'{e} J.-M. 1998, MNRAS,
   298, 1048
 \bibitem[]{}
   Hayashida K., Inoue H., Koyama K., Awaki H., Takano S., 1989, PASJ, 41, 373
 \bibitem[]{}
   Hynes R. I. et al. 1998, MNRAS, 300, 64
 \bibitem[]{}
   Inoue H. 1993, in Accretion Disks in Compact Stellar Systems, ed. 
                 J. C. Wheeler
 \bibitem[]{}
   King A. R. 1998, MNRAS, 296, L45
 \bibitem[]{}
   King A. R., Frank J., Kolb U., Ritter H. 1997a, ApJ, 484, 844
 \bibitem[]{}
   King A. R., Kolb U., Burdieri L, 1996, ApJ, 464, L127
 \bibitem[]{}
   King A. R., Kolb U., Szuszkiewicz E., 1997b, ApJ, 488, 89
 \bibitem[]{}
   King A. R., Ritter H. 1998, MNRAS, 293, L42
 \bibitem[]{}
   Lightman A. P., Zdziarski A. A. 1987, ApJ, 319, 643
 \bibitem[]{}
   Magdziarz P., Zdziarski A. A., 1995, MNRAS, 273, 837
 \bibitem[]{}
   Menou K., Narayan R., Lasota J.-P. 1999, ApJ, 513, 811
 \bibitem[]{}
   Mitsuda K. et al., 1984, PASJ, 36, 741
 \bibitem[]{}
   Miyamoto S., Kitamoto S., Iga S., Negoro H., Terada K. 1992, ApJ, 391, L21
 \bibitem[]{}
   Morrison R., McCammon D., 1983, ApJ, 270, 119
 \bibitem[]{}
   Oosterbroek T. et al.\ 1996, A\&A, 309, 781
 \bibitem[]{}
   Oosterbroek T. et al.\ 1997, A\&A, 321, 776
 \bibitem[]{}
   Poutanen J., 1998, in Abramowicz M. A., Bj\"{o}rnsson G., Pringle J. E., 
   eds, Theory of Black Hole Accretion Discs, CUP, Cambridge, in press,
   (astro-ph/9805025)
 \bibitem[]{}
   Pozdnyakov L. A., Sobol I. M., Sunyaev R. A. 1983, Ap. Space Phys. Rev.,
        2, 189
 \bibitem[]{}
   Ross R. R., Fabian A. C., Young A. J. 1998, MNRAS, in press
 \bibitem[]{}
   R\'{o}\.{z}a\'{n}ska A., Czerny B., \.{Z}ycki P., Pojma\'{n}ski G. 1999,
   MNRAS, in press
 \bibitem[]{}
   Rybicki G. B., Lightman A. P. 1979, Radiative Processes in Astrophysics.
   John Wiley \& Sons, New York
 \bibitem[]{}
   Shahbaz T., Ringwald F. A., Bunn J. C., Naylor T., Charles P. A.,
   Casares J., 1994, MNRAS, 271, L10
 \bibitem[]{}
   Shahbaz T., Charles P. A., King A. R. 1998, MNRAS, 301, 382
 \bibitem[]{}
   Shakura N. I., Sunyaev R. A. 1973, A\&A, 24, 337
 \bibitem[]{}
   Shimura T., Takahara F. 1995, ApJ, 445, 780
 \bibitem[]{}
   Skibo J. G., Dermer C. D., Ramaty R., McKinley J. M. 1995, ApJ, 446, 86
 \bibitem[]{}
   Smak J. I. 1984, Acta Astron., 34, 161
 \bibitem[]{}
   Sunyaev R. A., Titarchuk L. N. 1980, A\&A, 86, 121
 \bibitem[]{}
   Titarchuk L. 1994, ApJ, 434, 570
 \bibitem[]{}
   Takizawa M. et al.\ 1997, ApJ, 489, 272
 \bibitem[]{}
   Tanaka Y.,  Lewin W. H. G.  1995, in X--Ray Binaries, ed.
   W. H. G. Lewin, J. van Paradijs \& E. van den Heuvel (Cambridge: Cambridge
    Univ. Press), 126
 \bibitem[]{}
   Tanaka Y., Shibazaki N. 1996, ARA\&A, 34, 607
 \bibitem[]{}
   Turner M. et al.\ 1989, PASJ, 41, 345
 \bibitem[]{}
    van Paradijs J. 1996, ApJ, 464, L139
 \bibitem[]{}
   van der Klis M. 1995, in X--Ray Binaries, ed.
   W. H. G. Lewin, J. van Paradijs \& E. van den Heuvel (Cambridge: Cambridge
    Univ. Press), 252
 \bibitem[]{}
    \.{Z}ycki P. T., Czerny B., 1994, MNRAS, 266, 653
 \bibitem[]{}
    \.{Z}ycki P. T., Done C.,  Smith D. A. 1998, ApJ, 496, L25 
 \bibitem[]{}
    \.{Z}ycki P. T., Done C.,  Smith D. A. 1999, MNRAS, 305, 231 (Paper I)

\label{lastpage}

\end{thebibliography}
\end{document}